\documentclass[aps,prl,twocolumn,groupedaddress,longbibliography]{revtex4-1}
\usepackage{amsfonts,amsmath}
\usepackage{graphicx}
\usepackage[none]{hyphenat}

\usepackage{amsmath}
\usepackage{bm}
\usepackage{float}
\usepackage{color}
\usepackage{sidecap}
\allowdisplaybreaks
\usepackage{soul}
\setstcolor{red}
\usepackage{hyperref}

\begin{document}

\graphicspath{{figures/}}

\title{Prethermal memory loss in interacting quantum systems coupled to thermal baths}

\author{Ling-Na Wu}
\email[]{lnwu@pks.mpg.de}
\affiliation{Max Planck Institute for the Physics of Complex Systems, N\"othnitzer Str.~38,
D-01187, Dresden, Germany }

\author{Andr{\'e} Eckardt}
\email[]{eckardt@pks.mpg.de}
\affiliation{Max Planck Institute for the Physics of Complex Systems, N\"othnitzer Str.~38,
D-01187, Dresden, Germany }

\date{\today}

\begin{abstract}
We study the relaxation dynamics of  an extended Fermi-Hubbard
chain with a strong Wannier-Stark potential tilt coupled to a bath. When the system is subjected to
dephasing noise, starting from a pure initial state the system's total von Neumann entropy is found to
grow monotonously. The scenario becomes rather different when the system is
coupled to a thermal bath of finite temperature. Here,
for sufficiently large field gradients and initial energies, the
entropy peaks in time and almost reaches its largest possible value
(corresponding to the maximally mixed state), long before the system relaxes to
thermal equilibrium. This entropy peak signals an effective prethermal memory loss and,
relative to the time where it occurs, the system is found to exhibit a simple scaling behavior in space and time. By comparing the system's dynamics to that of a
simplified model, the underlying mechanism is found to be related to the
localization property of the Wannier-Stark system, which favors dissipative coupling
between eigenstates that are close in energy.

%

\end{abstract}

\maketitle


The problem of particles moving in a tilted lattice~(Wannier-Stark system)
is associated with various interesting phenomena, such as
Bloch oscillations~\cite{bloch1929quantenmechanik}, Stark localization~\cite{PhysRev.117.432} and Landau-Zener tunneling~\cite{zener1934theory}.
Despite its long history, it has kept to be a frontier research topic both theoretically~\cite{gluck2002wannier} and experimentally ~\cite{PhysRevLett.60.2426,PhysRevLett.83.4752,PhysRevLett.83.4756,PhysRevLett.76.4512,PhysRevLett.76.4508,PhysRevLett.87.140402,
PhysRevLett.105.215301,PhysRevLett.96.053903,PhysRevLett.102.076802,Mukherjee_2015,Schmidt2018,guardadosanchez2019subdiffusion}.
Recent studies~\cite{van2018bloch,schulz2018stark} show that an interacting Wannier-Stark system can exhibit non-ergodic behavior analogous to disorder-induced many-body localization~(MBL)~\cite{Altman2015MBLReview, Nandkishore2015MBLReview, ALET2018MBLReview,RevModPhys.91.021001}.
Understanding both the differences and similarities between such disorder-free and conventional MBL constitutes a fundamental question, which currently attracts a lot of attention~\cite{khemani2019localization,PhysRevLett.123.016601,taylor2019experimental,PhysRevLett.120.030601,PhysRevLett.119.176601,Grover2014JSM,PhysRevB.91.184202,
PhysRevLett.117.240601,PAPIC2015714,PhysRevLett.118.266601,PhysRevLett.119.176601,DeRoeck2014,Hickey2016JSM,PhysRevB.92.100305,carleo2012localization}.

On the other hand, there is an increased recent interest in the non-equilibrium properties of \emph{open} many-body quantum systems~\cite{BreuerPetruccione,carmichael2013statistical,Zoller2010PhysRevA.82.063605,daley2014quantum,PhysRevLett.104.160601,diehl2008quantum,
verstraete2009quantum,RevModPhys.89.015001,DanielPRL,DanielPRE,PhysRevLett.119.140602,PhysRevLett.109.045302,PhysRevLett.120.020401,baumann2010dicke,RevModPhys.85.553,PhysRevLett.111.073603,
PhysRevLett.121.170402,nakagawa2019negativetemperature,PhysRevLett.107.140404,PhysRevLett.116.235302,PhysRevX.7.011034,
Wu2019NJP,Wu2019PhysRevLett.123.030602,Levi2016PhysRevLett.116.237203,Fischer2016PhysRevLett.116.160401,
Medvedyeva2016PhysRevB.93.094205,Everest2017PhysRevB.95.024310,Marko2016AP,PhysRevLett.108.233603,PhysRevLett.113.210401,PhysRevLett.123.090402,PhysRevA.99.012106,Wang2014}.
While the coupling to an environment of finite temperature constitutes a natural situation, it is rather cumbersome to simulate. Therefore, the impact of dissipation is often treated by using dephasing noise~\cite{gardiner2004quantum} as a simpler bath model. Even though dephasing noise will eventually drive the system into an infinite-temperature state, it is assumed to qualitatively capture the effect of weak coupling to a bath on the transient evolution. Understanding, under which circumstances this assumption breaks down is an important question for the simulation of open many-body systems~(see also Refs.~\cite{PhysRevA.97.053610,tan2019interactionimpeded}).

Here we report on a surprising phenomenon in the relaxation dynamics of an interacting Wannier-Stark system coupled to a thermal bath of finite temperature. It sheds light on both of the above questions, since it can neither be observed for a disorder-localized system nor for dephasing noise. In a large parameter regime, we find that long before the system reaches thermal equilibrium, it transiently approaches the maximally mixed state. This effect implies an effective prethermal memory loss.
It is reminiscent of the
universal dynamics recently observed in
isolated quantum gases \cite{2018Natur.563..217P,2018Natur.563..225E}.

We consider a one-dimensional extended Hubbard chain half
filled with spin-polarized fermions and subjected to a linear potential
gradient. It is described by the Hamiltonian
\begin{equation}\label{Ham}
\resizebox{.9\hsize}{!}{$H = -J\sum\limits_{i=1}^{M-1}{\left(c_i^\dag c_{i+1} + c_{i+1}^\dag c_i \right)} + \sum\limits_{i=1}^{M}{W_i n_i}+V\sum\limits_{i=1}^{M-1}{n_i n_{i+1}}.$}
\end{equation}
Here $c_i^\dag$ and $n_i=c^\dag_ic_i$ are the creation and number operator for a fermion on lattice site $i$.
Moreover, $J$ is the
tunneling parameter, $W_i=-r i$ captures the potential gradient $r$, and $V$
quantifies nearest-neighbor interactions. The single-particle (bulk)
eigenstates, known as Wannier-Stark states, are centered at the lattice
sites, with a localization length $\sim J/2r$ (with respect to the site
index~$i$) and energies that increase by $r$ from site to site{~\cite{PhysRev.117.432}}.
Moreover, also the interacting system shows properties akin to MBL~\cite{van2018bloch,schulz2018stark}. Henceforth, we use $J$, $J/k_B$, and
$\hbar/J$ as units for energy, temperature, and time, respectively, so that
$J=\hbar=k_B=1$.

\begin{figure}[!htbp]
    \centering
{\includegraphics[width=1\columnwidth]{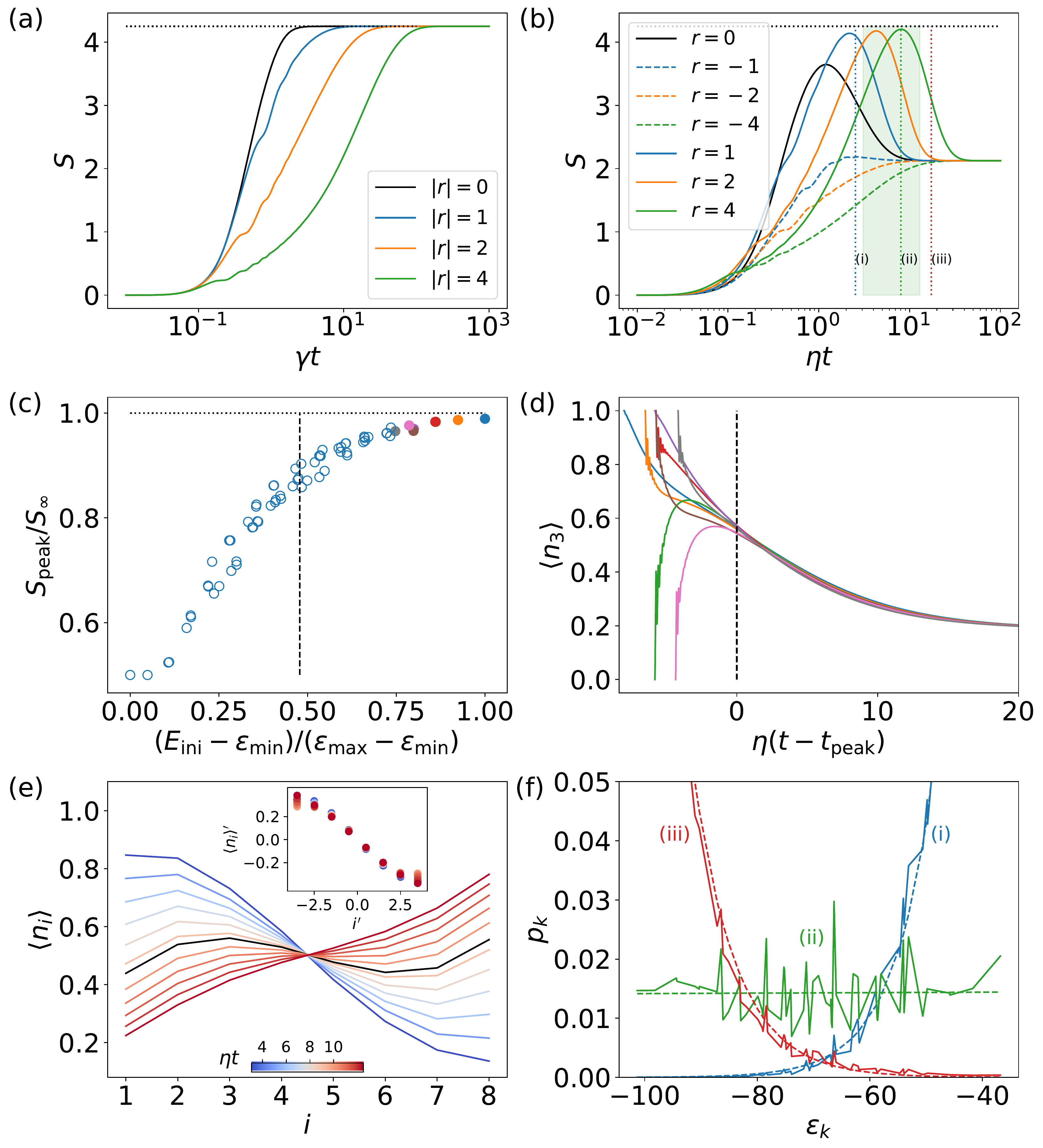}}
    \caption{(a), (b) Time evolution  of the  von Neumann entropy for the tilted ladder (\ref{Ham}) initialized in a Fock state with the left half of the chain occupied and coupled to (a) a dephasing bath [Eq.~\eqref{L}] or (b) a thermal bath [Eq.~\eqref{heat}].
    The dotted line marks the largest possible entropy $S_{\infty}$.
    (c)~Normalized peak entropy for initial Fock states of various energies
		$E_\text{ini}$. The vertical dashed line marks $E_{\infty}$.
    (d)~Evolution of the mean occupation of the third lattice site relative to the time $t_\text{peak}$, where the entropy peak is reached. The line colors mark different initial Fock states corresponding to colored bullets in (c).
    (e)~Density profile $\langle n_i\rangle$ at equidistant times during the
		time window marked by the shaded area in (b). The black line marks
		$t=t_\text{peak}$. The inset shows the collapse of all the curves by
    rotating them by an angle proportional to the corresponding time.
    (f)~Diagonal elements of the density matrix $p_k$ at three points in time
		[marked by (i)-(iii) in (b)] (solid lines) compared to effective thermal states of identical average energy (dashed lines).
    The parameters are $M=8$, $V=1$, $\gamma=\eta=0.1$, and $T$ so that
		$S_T=S_\infty/2$. The field gradient for (c)-(f) is $r=4$.
		}\label{HF0}
\end{figure}

When coupled weakly to a thermal bath, which is modeled as a collection of harmonic
oscillators in thermal equilibrium and couples to the on-site occupations, the system
can be described by a Redfield master equation~\cite{BreuerPetruccione},
    \begin{align} \label{heat}
    \frac{d \rho}{dt} =  & -i\left[H,\rho\right]
			 +  \eta \sum\limits_{k,q,p,l=1}^{M}{} \left[{R_{kqlp}\left( L_{kq}\rho L_{pl}^\dag- L_{pl}^\dag L_{kq} \rho \right)} \right. \notag\\
			& +  \left. {R_{plqk}\left( L_{kq}\rho L_{pl}^\dag -\rho  L_{pl}^\dag L_{kq} \right)}\right],
    \end{align}
with jump operators $L_{kq} = |k\rangle \langle q|$ between many-body eigenstates $|k\rangle$ of energy $\varepsilon_k$~\cite{DanielPRE,Wu2019NJP}.
The corresponding transition rates read~$R_{kqpl} = \pi v_{kqpl} g({\varepsilon}_k - {\varepsilon}_q)$, with
 $v_{kqpl} = \sum\limits_{i=1}^M{\langle k|n_i |q\rangle \langle p|n_i |l\rangle}$ and bath correlation function $g(E) = J(E)/(e^{E/T}-1)$, where we assume an
Ohmic spectral density $J(E) = E$.

In the high-temperature limit, we have $g(E) \simeq T$ and the
transition rate $R_{kqpl}$ becomes independent of energy. Thus, the master equation reduces to
\begin{equation}\label{L}
\frac{d \rho}{dt} = -i\left[H,\rho\right]
	+ \gamma \sum\limits_{i=1}^{M}{\left(n_i \rho n_i
		- \frac{1}{2}n_i^2\rho -  \frac{1}{2}\rho n_i^2\right)},
\end{equation}
with $\gamma = \eta T$, describing dephasing noise.

Figure~\ref{HF0} shows the time evolution of the von Neumann entropy
$S=-\mathrm{tr}\{\rho\log(\rho)\}$ of the total system, when coupled to (a)
a dephasing bath or (b) a finite-temperature bath. It is calculated by
numerically integrating Eqs.~\eqref{L} and~\eqref{heat},
respectively, starting from a Fock state with the left half of the chain
occupied. The temperature $T$ of the thermal bath is chosen such that the
corresponding equilibrium entropy, $S_T$~(obtained for the Gibbs state
$\rho_T={\cal Z}_T^{-1}\sum_k e^{-\varepsilon_k/T}|k\rangle\langle k|$ with
${\cal Z}_T=\sum_k e^{-\varepsilon_k/T}$), is equal to half the largest
possible entropy $S_{\infty} \equiv S_{T=\infty}= \log({\cal D})$, with Hilbert
space dimension ${\cal D}=M!/[(M/2)!]^2$~(see Fig.~\ref{Tvec} of Ref.~\cite{sm} for other temperatures).
For dephasing noise~[Fig.~\ref{HF0}(a)], the entropy grows monotonously
to the maximum value $S_{\infty}$, being insensitive to the sign of the
potential gradient $r$. In turn, when the system is coupled to the finite-
temperature bath~[Fig.~\ref{HF0}(b)], the entropy approaches its equilibrium
value rather differently for negative and positive $r$. While in the former case
the entropy grows monotonously to $S_T$ (except for small $|r|$), in the
latter case it first reaches a peak value $S_\text{peak}$ well above $S_T$,
before relaxing to equilibrium. As will become apparent in the following,
this difference can be attributed to the different mean energies of the
initial states.

Remarkably, we can observe in Fig.~\ref{HF0}(b) that for large positive
gradients $r$, the peak entropy almost reaches the largest possible entropy
$S_\infty$ (dotted line), which uniquely corresponds to the maximally mixed
state $\rho_\infty\equiv\rho_{T=\infty}=\mathcal{D}^{-1}\sum_k|k\rangle\langle k|$. This effect
can be observed for a wide range of initial conditions: In Fig.~\ref{HF0}(c)
we plot the peak entropies $S_\text{peak}$ reached during the evolution
starting from various initial Fock states, versus their mean energy
$E_\text{ini}$ (scaled between 0, for the ground-state energy $\varepsilon_\text{min}$,
and 1, for the energy $\varepsilon_\text{max}$ of the most excited state). Peak
entropies close to $S_\infty$ are found as long as $E_\text{ini}$ lies well
above the energy $E_\infty=\mathrm{tr}\{\rho_\infty H\}$ of the maximally mixed state (dashed line).

Since the maximally mixed state is unique, reaching an entropy peak with
$S_\text{peak}\approx S_\infty$ indicates that
we can expect the system dynamics to become (approximately) independent of the initial conditions near and
after approaching the peak entropy. Such a behavior is confirmed in Fig.~\ref{HF0}(d),
where we plot the evolution of the site occupation $\langle n_{i=3}\rangle$
relative to the time $t_\text{peak}$ at which the entropy peak is reached.
The different curves, which correspond to different initial states [labeled
by line colors corresponding to the colored bullets in Fig.~\ref{HF0}(c)], clearly
converge near $\eta(t-t_\text{peak})=0$ and subsequently show almost
identical behavior. Similar behavior can also be observed
for other site occupations $\langle n_i\rangle$ and for larger systems~(see Figs.~\ref{nt},~\ref{ntv} and Figs.~\ref{n_rate_eq},~\ref{nmf} of Ref.~\cite{sm}, respectively).
Thus, the system undergoes an effective \footnote{``Effective'', since formally the initial sate can still be recovered by evolving backwards in time.} prethermal memory loss, long before it reaches thermal equilibrium.

Moreover, we also find that the way the system approaches the maximally mixed
state shows a simple form of scaling behavior. In Fig.~\ref{HF0}(e) we plot the density
distribution, $\langle n_i\rangle$, at various times near
$t_\text{peak}$ [within the shaded area in Fig.~\ref{HF0}(b)], for $r=4$ and
starting from the Fock state with the left half of the chain occupied
[green curve in Fig.~\ref{HF0}(b)]. These density profiles
collapse on top of each other when rotated by an angle
proportional to the evolved time (see inset). Note that for $r=4$ the
Wannier-Stark states are already well localized on single lattice sites, so that
the plotted density profile (which can directly be measured in quantum-gas
systems) approximately corresponds to the occupation of the single-particle
eigenstates. In this sense, the observed behavior is somewhat reminiscent of the
universal scaling behavior recently observed in the occupations of
long-wavelengths momentum modes during the far-from equilibrium dynamics of
isolated quantum gases \cite{2018Natur.563..217P,2018Natur.563..225E}, which
was associated with the presence of non-thermal fixed points{~\cite{PhysRevLett.101.041603,PhysRevD.92.025041,PhysRevB.84.020506}}.

Further insight on how the system approaches the maximally mixed state is gained
by looking at the probability distribution $p_k=\langle k|\rho|k\rangle$ for
occupying many-body energy eigenstates $|k\rangle$. In Fig.~\ref{HF0}(f) we
plot the distribution $p_k$ (solid lines) for $r=4$ at three times: (i)
slightly before, (ii) at, and (iii) slightly after $t_\text{peak}$ [as
indicated in Fig.~\ref{HF0}(b) relative to the green curve]. Interestingly, these distributions agree
rather well to those for thermal states $\rho_{T_\text{eff}}$ (dashed lines)
with the effective temperature $T_\text{eff}$ determined by the instantaneous
energy, $\mathrm{tr}\{\rho_{T_\text{eff}} H\}\equiv E=\mathrm{tr}\{\rho H\}$.
This observation suggests that the prethermal memory loss is due to a \emph{dissipative form of
prethermalization}, where the system rapidly approaches a Gibbs state, whose
effective temperature $T_\text{eff}$ then slowly relaxes to the equilibrium
temperature $T$. This scenario immediately explains that an infinite
temperature state is approached long before the system has thermalized as long as the initial
energy $E_\text{ini}$ lies well above the infinite-temperature energy
$E_\infty$. Namely, in this case the system has enough time to approach a
prethermal state $\rho_{T_\text{eff}}$, before $1/T_\text{eff}$ passes through zero
from below at the time when $E$ drops below $E_{\infty}$.

The prethermal relaxation to a Gibbs-like state with slowly varying effective temperature occurs in a way rather different from standard (pre)thermalization~\cite{PhysRevLett.93.142002}. It cannot be understood as the prethermalization of system and bath together, since the bath always remains in the same thermal state with constant temperature $T$. Nor can it be explained by the thermalization of the system itself on a time scale that is fast compared to slow energy dissipation by the bath, since the system is localized and non-ergodic. Let us, therefore, investigate the underlying mechanism and its relation to Stark localization.

\begin{figure}[!htbp]
    \centering
    {\includegraphics[width=0.9\columnwidth]{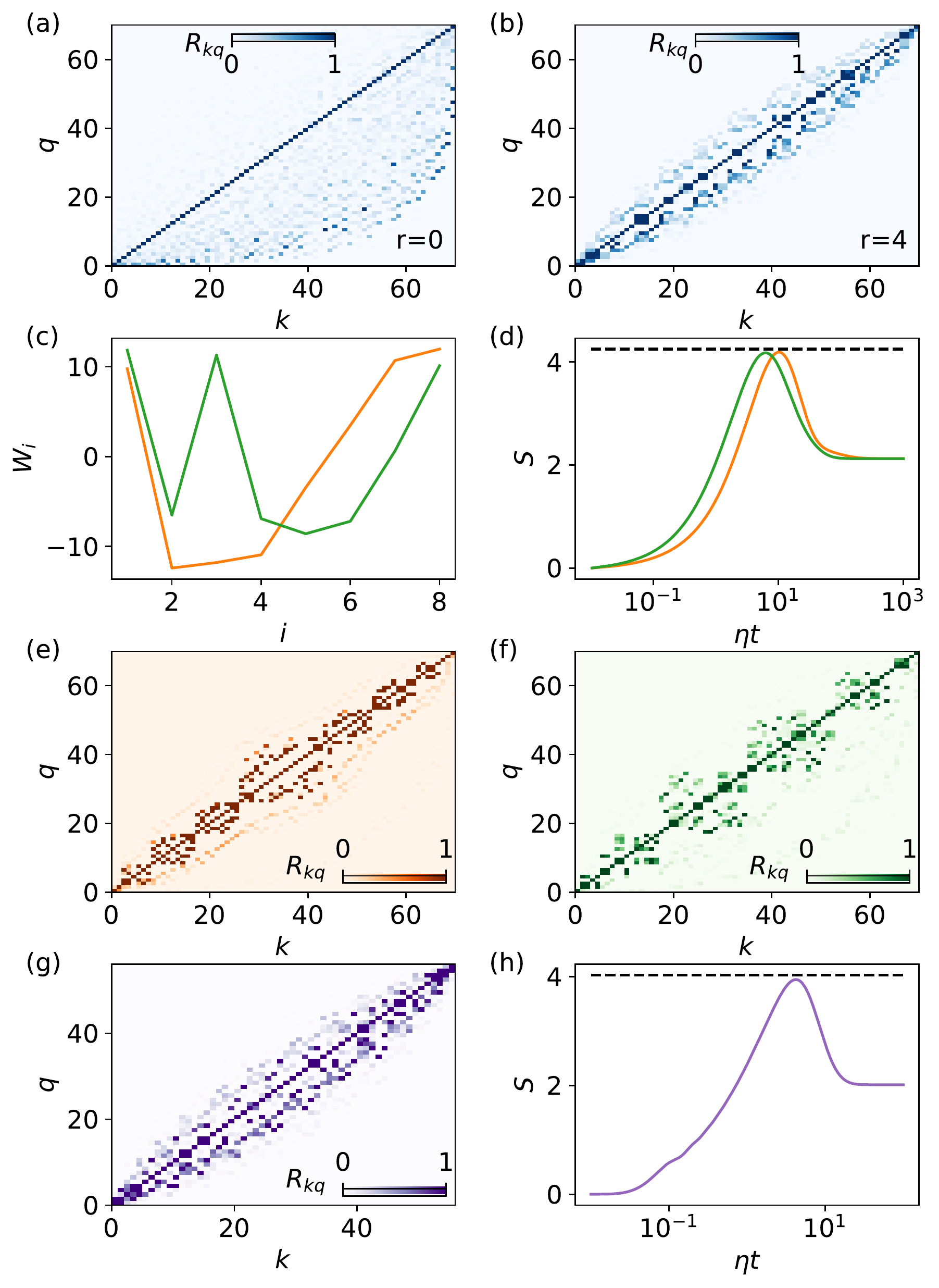}}
    \caption{(a), (b) Rate matrix for the Fermi-Hubbard chain coupled to a thermal bath, with $r=0$ (a) and $r=4$ (b).
    (c) Examples for on-site potentials $W_i$ that were optimized over the range $[-16,16]$ to provide a large peak entropy when evolving from the most excited state. The results are representative for $500$ numerical runs.
     (d) Time evolution of the corresponding entropies. 
     (e)-(f) Corresponding rate matrices $R_{kq}$.
    (g) Rate matrix and (h) entropy for a tilted bosonic Hubbard chain with $r=4$ and initially all particles occupying the leftmost site.
    The system sizes are $M=8$, $N=4$ for (a)-(f), and $M=6$, $N=3$ for (g)-(h). Other parameters are
		$V=1$, $\eta=0.1$, and $T$ so that $S_T=S_\infty/2$.
    		}\label{rate2}
\end{figure}

Although a two-level system prepared in its excited state passes through the maximum
entropy state while equilibrating with a thermal bath~\cite{sm}, such behavior is highly nontrivial for interacting many-body systems.
To figure out the conditions for a close-to-maximum peak entropy, let us
focus on the weak-coupling limit, where the secular approximation~\cite{
BreuerPetruccione,carmichael2013statistical} gives a Lindblad master equation,
\begin{equation*}
 \frac{d \rho}{dt} =  -i\left[H,\rho\right] + \eta \sum\limits_{k,q=1}^{M}{} R_{kq}{\left( L_{kq}\rho L_{kq}^\dag- \frac{1}{2}\left\{L_{kq}^\dag L_{kq}, \rho\right\} \right )},
\end{equation*}
with rates $R_{kq} \equiv R_{kqqk}$. They obey $R_{kq}/R_{qk}=e^{-(\varepsilon_k-\varepsilon_q)/T}$,
so that low energy states are favored and the system is asymptotically driven towards the Gibbs state $\rho_T$. The matrix elements
$\rho_{kq} \equiv \langle k | \rho |q \rangle$ follow
$\dot \rho_{kq} = -i ({\varepsilon}_k -{\varepsilon}_q)\rho_{kq}
+ \eta  \sum_p{\left[R_{kp}\rho_{pp}\delta_{kq}-\frac{1}{2} (R_{pk}+R_{pq})\rho_{kq} \right]}$, where diagonal and off-diagonal elements
decouple from each other. The latter decay with rates
$\Gamma_{kq}=\frac{1}{2} \eta  \sum_p{ (R_{pk}+R_{pq})}$ and have to be
negligible already when the peak entropy is reached, to allow for the observed
transient approach of the maximally mixed state~(this is indeed the case, see Fig.~\ref{init} and discussion in Ref.~\cite{sm}). The dynamics of the diagonal
elements $p_k \equiv \rho_{kk}$ is determined by the rate matrix $R_{kq}$
through the Pauli rate equation $\dot p_k =  \eta \sum_q{\left[R_{kq}p_q-R_{qk}p_k \right]}$ \cite{BreuerPetruccione}.

In Fig.~\ref{rate2}~(a) and (b) we compare rate matrices for $r=0$ and $r=4$.
Without potential gradient, one finds long-range coupling with respect to
energy. In contrast, at a large field gradient ($r=4$) the transition rates
predominantly couple states that are close by in energy. The latter is a
consequence of Stark localization, where eigenstates that are close in space,
so that they are coupled by the bath via the densities $n_i$, are close also with
respect to energy.
Note that for a disorder-localized Fermi-Hubbard chain without this spatio-energetic correlation we find non-local rate matrices and no prethermal memory loss (see Fig.~\ref{disorder} of Ref.~\cite{sm}). This constitutes an even more drastic difference between the relaxation dynamics of Stark and disorder-induced MBL than the one observed for dephasing noise~\cite{Wu2019PhysRevLett.123.030602}.

To check, whether a rate matrix with energy-local coupling is crucial for the
appearance of a close-to-maximum entropy ($S_{\rm peak} \simeq S_{\infty}$),
we investigate the rate matrices for rather different on-site
potentials $W_i$~[see Fig.~\ref{rate2}(c)] that (were optimized to) equally
give rise to large peak entropies [see Fig.~\ref{rate2}(d)]. It turns out
that, indeed, they also show near-neighbor coupling
[see Figs.~\ref{rate2}(e)-(f)].
Moreover, also the tilted \emph{bosonic} Hubbard chain [given by Eq.~(1) with bosonic annihilation operators $c_i$ and the last term replaced by on-site interactions $\frac{1}{2}V\sum_i n_i(n_i-1)$] shows $S_\text{peak}\approx S_\infty$ together with an energy local rate matrix [see Figs.~\ref{rate2}(g)-(h)].
This, together with results for simplified models shown below (and in Fig.~\ref{rate_texture} of Ref.~\cite{sm}) indicates clearly that the prethermal memory loss discussed here is a very robust phenomenon.

To address the question, why the
appearance of a close-to-maximum entropy is associated with local coupling
between energy states, let us now consider a simplified rate model. Here the
energy eigenstates have equally spaced non-degenerate
energies $\varepsilon_k=rk$, with $k=0,1,\cdots \mathcal{D}-1$,
and are coupled by thermal rates that are homogeneous and local with respect
to energy. We define $R_{k+n,k}\equiv R_n$, where $R_n=0$ for $|n|>n_m$ and
$R_n=g(nr)$ for $|n|\le n_m$, so that (as a property of the bath correlation
function $g$) $R_n/R_{-n}=(R_+/R_{-})^n=e^{-nr/T}$, with
$R_\pm\equiv R_{\pm1}$.

\begin{figure}
  \centering
    \includegraphics[width=0.95\columnwidth]{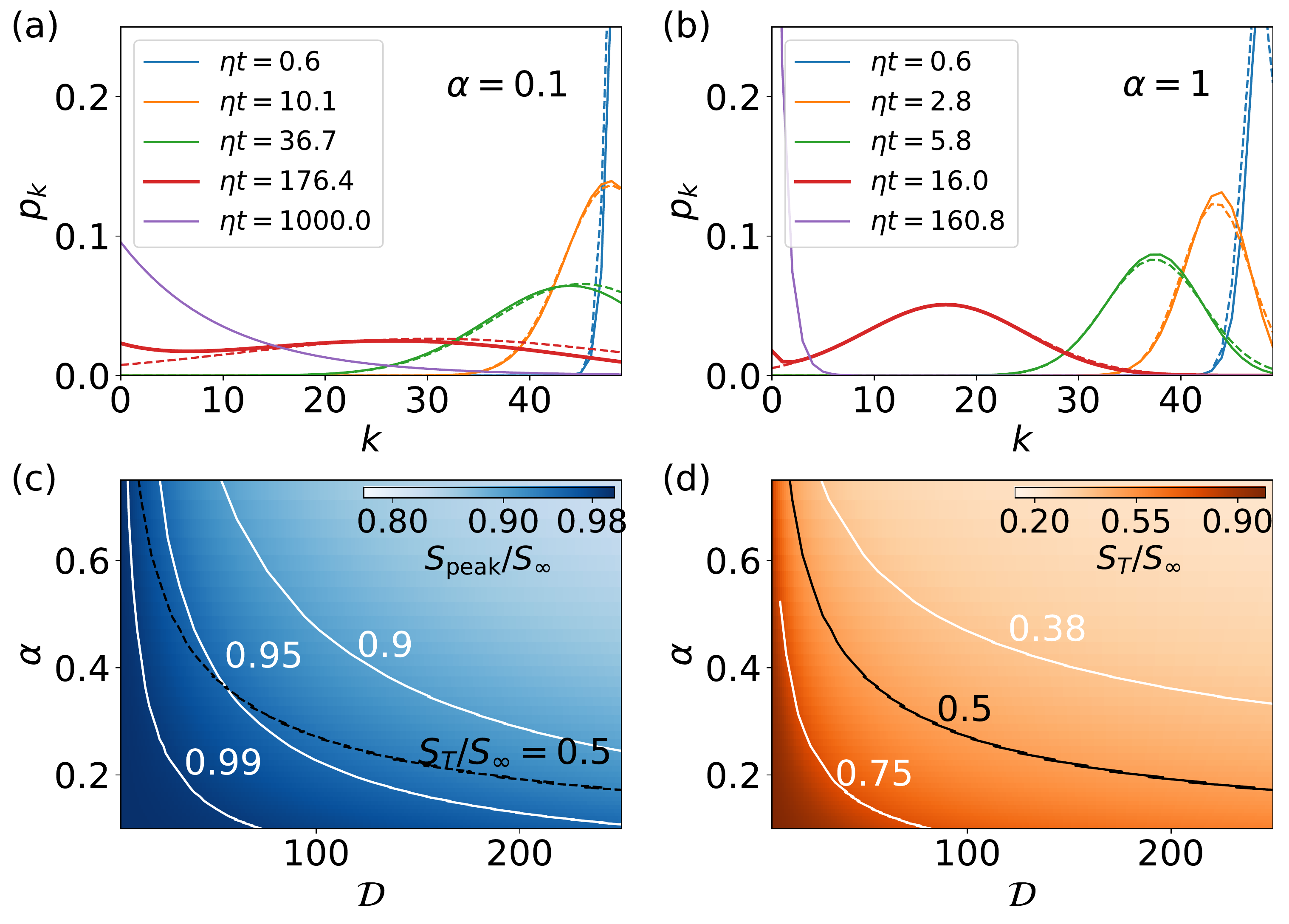}
  \caption{
	Simplified rate model with nearest-neighbor coupling $n_m=1$ evolving from most escited state. 	
	(a), (b) Probability distribution $p_k$ at different times for $\mathcal{D}=50$ and $\alpha=0.1$ (a) and 1 (b).
	(c), (d) Normalized peak and thermal entropies $S_\text{peak}/S_\infty$ (c) and $S_{\rm T}/S_\infty$ (d) versus $\alpha$ and ${\cal D}$.
}\label{DD}
\end{figure}

Let us first study nearest-neighbor coupling, $n_m=1$. By defining $\nabla^2 p_k =p_{k+1}+p_{k-1}-2p_k$ and
$\nabla p_k=(p_{k+1}-p_{k-1})/2$, we can write the Pauli rate equation as a discrete drift-diffusion equation,
$\dot p_k = \eta(\bar R \nabla^2 p_k + \delta R \nabla p_k)$~\cite{RevModPhys.15.1},
where diffusion and drift are quantified by $\bar R \equiv (R_+ + R_-)/2$ and
$\delta R \equiv R_- - R_+$, respectively. Scaling time with $(\eta \bar R)^{-1}$, the
model is completely characterized by the ratio $\alpha=\delta R/{\bar R}$ and
the system size $\mathcal{D}$.
Starting from the highest excited state, $p_k(0)=\delta_{k,k_\text{ini}}$ with
$k_\text{ini}=\mathcal{D}-1$, and setting $\mathcal{D}=50$, in Figs.~\ref{DD}(a)
and (b) we plot $p_k$ for $\alpha=0.1$ and $\alpha=1$ for different times (solid lines, fat red lines indicate $t_\text{peak}$).
While for $\alpha=0.1$, a rather uniform
distribution is found at a time $t_\text{peak}$, approximating the maximum
entropy state with $p_k=1/\mathcal{D}$, this is not the case for larger drift,
$\alpha=1$. Before reaching thermal equilibrium, we find the distribution well
approximated by a Gaussian of standard deviation
$\sigma=\sqrt{2{\bar R} t}$ centered at $k_0=k_\text{ini}-\delta R t$ \cite{RevModPhys.15.1} (dashed lines).
The condition for reaching an almost flat
distribution is, thus, given by the intuitive requirement that the drift time
needed to reach $k_0=\mathcal{D}/2$, $\tau_F=(k_\text{ini}-\mathcal{D}/2)/\delta R$, is larger than the diffusion time giving rise to
$\sigma=\mathcal{D}/2$, $\tau_D=\mathcal{D}^2/(8 {\bar R})$. Thus, for
$k_\text{ini}=\mathcal{D}-1$ we expect $S_\text{peak}\approx S_\infty$
as long as $\alpha \lesssim 4\mathcal{D}^{-1}$, which is confirmed in
Fig.~\ref{DD}(c) [see also Fig.~\ref{Sam}(a) of Ref.~\cite{sm}].

However,
$S_\text{peak}\approx S_\infty$ is a non-trivial result only as long as the
thermal entropy $S_T$, plotted in Fig.~\ref{DD}(d), lies well below $S_\infty$.
{For $\mathcal{D}r \gg T$ we can neglect the upper bound of the spectrum and
$S_T$ approaches the value for an harmonic oscillator with frequency $r$~\cite{gould2010statistical},
$S_T \simeq \frac{x}{x-1}\log(x)-\log(1-x)$ with
 $x=e^{-r/T}=R_+/R_-=(2-\alpha)/(2+\alpha)$, so that for $\alpha\ll1$ one has
$S_T\simeq \log(1/\alpha)+1+\mathcal{O}(\alpha)$ [Fig.~\ref{DD}(d)]. Thus,
$S_T/S_\infty<s$ as long as $\alpha \gtrsim e\mathcal{D}^{-s}$. }While  for $\mathcal{D}\to\infty$ this
requirement is incompatible with the one for large peak entropies,
$\alpha \lesssim 4\mathcal{D}^{-1}$, { it turns out
that the different prefactors appearing in both conditions (whose values can deviate from our estimates $e$ and $4$) still give rise to a large non-trivial regime for finite
$\mathcal{D}$}, as can be inferred from Figs.~\ref{DD}(c) and (d).

\begin{figure}
  \centering
\includegraphics[width=0.494\columnwidth]{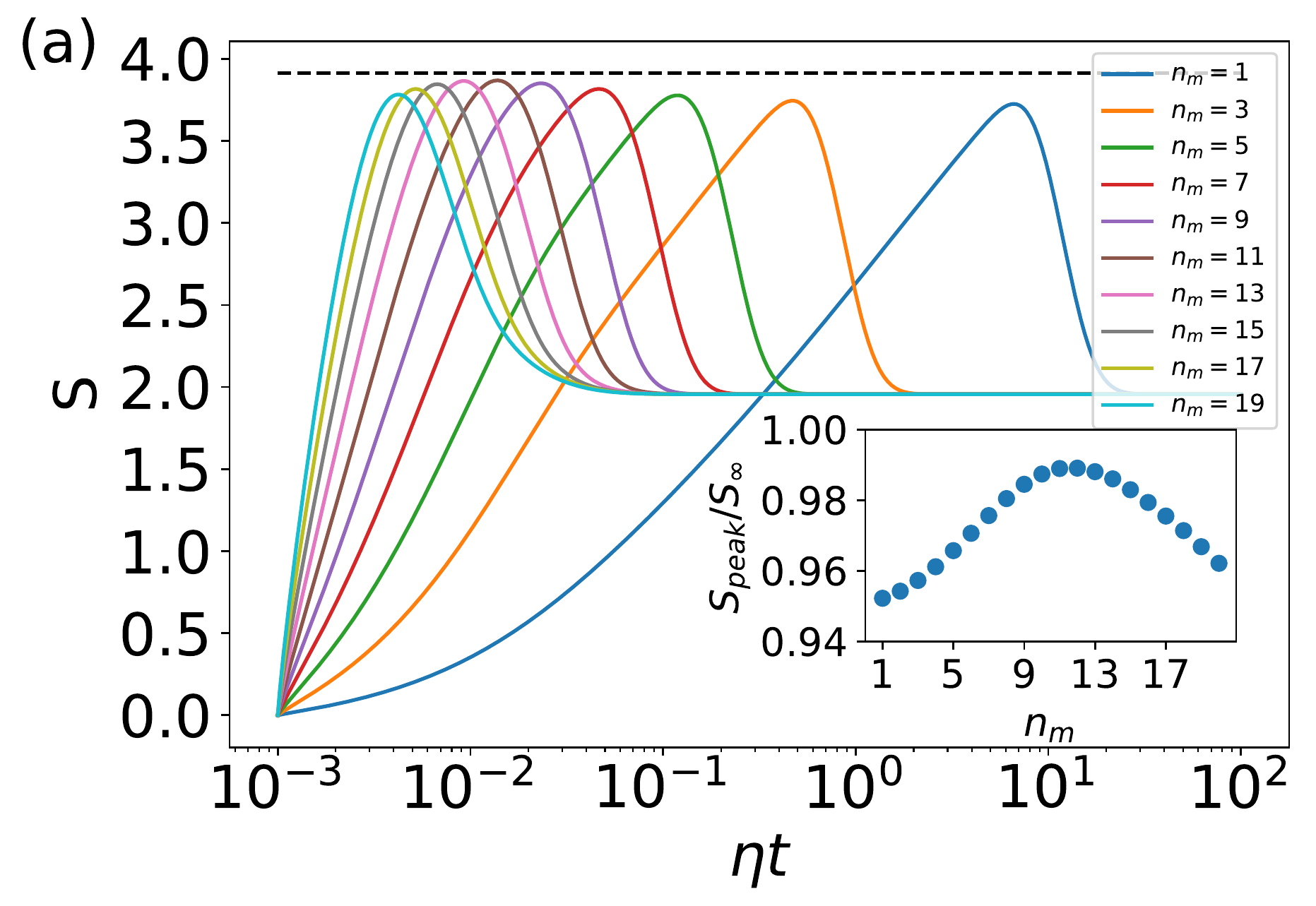}
  \includegraphics[width=0.494\columnwidth]{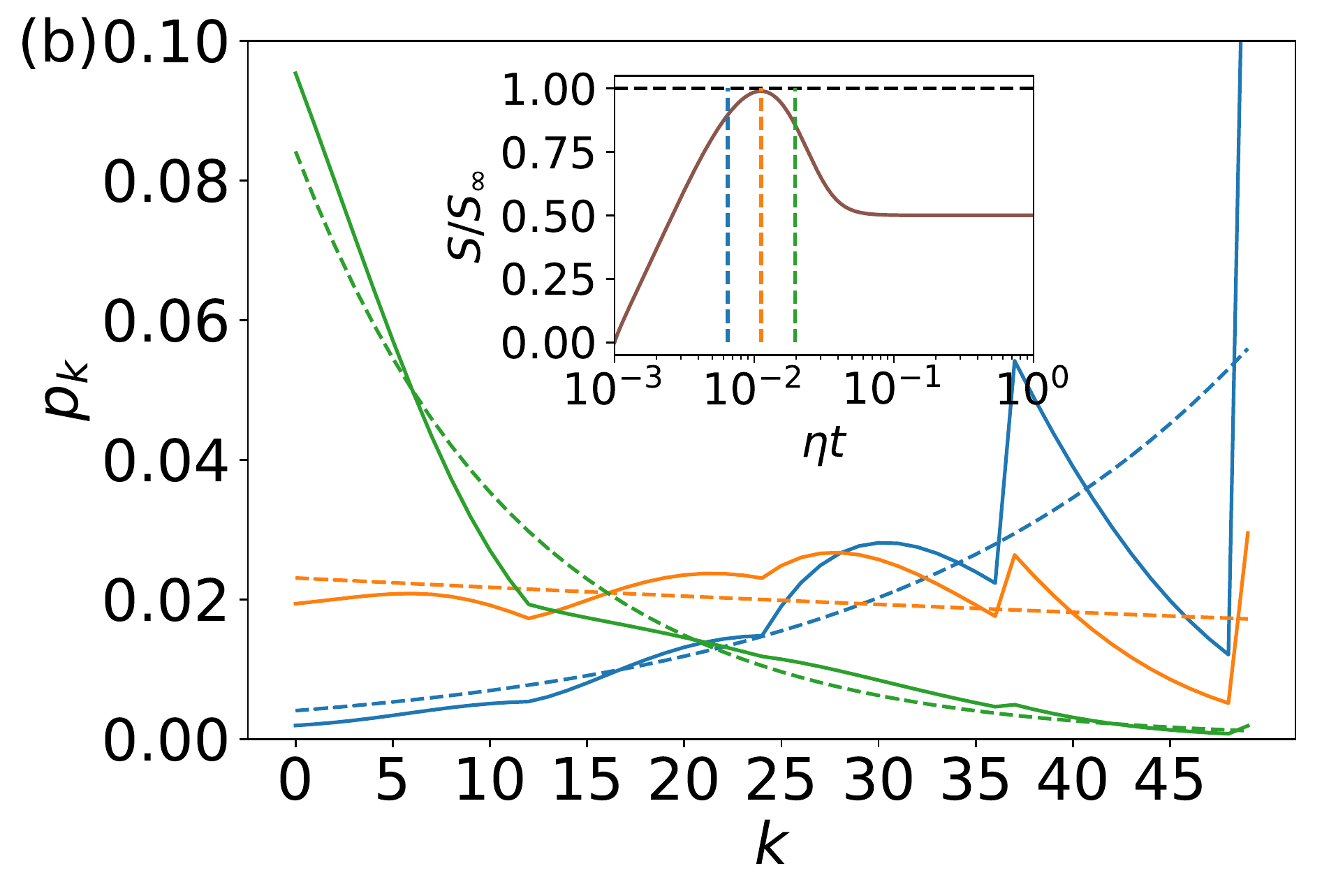}
	\caption{Evolution of the simplified rate model starting from the most excited state for $\mathcal{D}=50$, $\eta=0.1$, and $T$ so that $S_T=S_{\infty}/2$.
	(a) Entropy evolution for different $n_m$. Inset: Normalized peak entropies versus $n_m$.
	(b)~Distribution $p_k$ (solid lines) for $n_m=11$ at the times marked in the
	inset compared to effective Gibbs states (dashed lines).
		}\label{DDv}
\end{figure}

The simplified rate model with $n_m=1$ roughly corresponds to the case of a
single particle in a tilted lattice. While it describes peak entropies
$S_\text{peak}\approx S_\infty$ and prethermal memory loss, it gives rise to a Gaussian rather than an exponential prethermal distribution.
This suggests that the formation of a prethermal Gibbs state requires more
complex rate matrices as they are found for the many-particle case. In
Fig.~\ref{DDv} we investigate, what happens when increasing the coupling range
$n_m$, and thus the complexity, of the simplified rate model [using
$\mathcal{D}=50$ and $\eta=0.1$]. We observe that $S_\text{peak}$ first
increases with $n_m$ before, after reaching a maximum at $n_m=11$, it decreases
again. {The first increase with $n_m$ might be explained by the prethermal
distribution becoming more Gibbs like and thus flatter at $t=t_\text{peak}$ than the Gaussian (which always retains a finite $\sigma$). This is confirmed in Fig.~\ref{DDv}(b), where we plot the distribution $p_k$ at different times for $n_m=11$ and find rather good agreement
with an effective Gibbs state.} The subsequent decrease of $S_\text{peak}$ can,
in turn, be attributed to an increase of the drift velocity with $n_m$, which
is clearly visible also in Fig.~\ref{DDv}(a) and which reduces the time
available for reaching a prethermal distribution. This mechanism explains, why
large transient peak entropies $S_\text{peak}\approx S_\infty$, are found for rate matrices that are local in energy.

In conclusion, we have shown that the non-equilibrium relaxation dynamics of interacting Wannier-Stark ladders coupled to a finite-temperature environment can feature effective prethermal memory loss. The effect is found to rely on a dissipative form of prethermalization. In experiment with ultracold atoms, a thermal environment could be provided, for instance, by the coupling to a second atomic species~\cite{PhysRevLett.121.130403}.

\begin{acknowledgments}
We acknowledge discussions with Markus Oberthaler.
This research was funded by the Deutsche
Forschungsgemeinschaft (DFG) via the Research Unit
FOR 2414 under Project No. 277974659.
\end{acknowledgments}


%

\pagebreak

\clearpage

\onecolumngrid
\begin{center}
  \textbf{\large Supplemental Material for\\``Prethermal memory loss in interacting quantum systems coupled to thermal baths"}\\[.2cm]
  Ling-Na Wu and Andr{\'e} Eckardt\\[.1cm]
  {\itshape Max Planck Institute for the Physics of Complex Systems, N\"othnitzer Str.~38,
D-01187, Dresden, Germany}
\end{center}

\setcounter{equation}{0}
\setcounter{figure}{0}
\setcounter{table}{0}
\setcounter{page}{1}
\renewcommand{\theequation}{S\arabic{equation}}
\renewcommand{\thefigure}{S\arabic{figure}}
\renewcommand{\bibnumfmt}[1]{[S#1]}

\section*{A two-level system}

Consider a two-level system with eigenstates $|+\rangle$, $|-\rangle$
and the corresponding eigenenergies ${\varepsilon}_{\pm}$.
According to the Lindblad master equation introduced in the main text, the dynamics of the corresponding occupation probabilities $p_\pm$ is governed by the Pauli master equation
\begin{equation}
\dot{p}_{+} = \eta(R_{+-}p_{-} - R_{-+}p_+) = -\dot{p}_-.
\end{equation}
Its solution is given by
\begin{equation}\label{p+}
p_+ = \alpha - (\alpha - p_+(0))e^{-\eta R_0 t} = 1 - p_-,
\end{equation}
where $R_0 = R_{+-} + R_{-+}$ and $\alpha = R_{+-}/R_0$.
While $\alpha=1/2$ for dephasing noise, for a thermal bath of finite temperature we find $\alpha = \left( 1+e^{2\Omega/T}\right)^{-1}$, with $\Omega = {\varepsilon}_+ - {\varepsilon}_-$ being the energy splitting.

The off-diagonal terms of the density matrix
\begin{equation}\label{rho}
\rho =
\begin{pmatrix}
p_+ & \rho_{+-} \\
\rho_{-+}  & 1-p_+
\end{pmatrix},
\end{equation}
are described by
 \begin{equation}\label{p+-}
\rho_{\pm \mp} = e^{\mp i \Omega t - \eta R_s t/2}\rho_{\pm \mp}(0),
\end{equation}
with $R_s = \sum_{k,q=\pm} R_{kq}$.

To quantify the difference between thermal bath and dephasing noise, we investigate the purity. It is calculated as $f=\sum_k \lambda_k^2$, with $\lambda_k$ being the eigenvalues of the density matrix.
It shows similar behavior as the von Neumann entropy $S=-\sum_{k}{\lambda_{k}\log(\lambda_k)}$, but is easier to handle analytically.
For the two-level system, we can verify that the purity is given by
\begin{eqnarray}\label{f}
f &=& { \frac{1}{2} + 2|\rho_{+-}|^2 + 2\left(p_+-1/2\right)^2} \notag\\
&=&  A_1 e^{-\eta R_0 t} + A_2 e^{-2 \eta R_0 t} + B e^{-\eta R_s t} + C,
\end{eqnarray}
with
\begin{eqnarray}\label{A}
A_1 &=&  2 (1-2\alpha)(\alpha-p_+(0)) , \notag\\
A_2 &=& 2 (\alpha - p_+(0))^2, \notag\\
B &=& 2|\rho_{+-}(0)|^2, \notag\\
C &=& 2\alpha(\alpha-1)+1.
\end{eqnarray}
Except for $A_1$, all the other coefficients are non-negative.
For dephasing noise, we have $\alpha = 1/2$, thus $A_1 = 0$. Hence, the purity $f$ decays monotonously to its steady state value $	C=1/2$.
While for a finite-temperature bath, $A_1$ can be negative, depending on the parameters of the system and the initial state.
When $A_1<0$,  there will be a competition between the first term and the others,
leading to a non-monotonous behavior in the purity. An obvious example is when we start from the excited state.
Then the purity will degrade from its initial value $1$~(for a pure state with $p_+=1$) to the minimal value $1/2$~(uniquely corresponding to the maximally mixed state with $p_+=p_{-}=1/2$), and then
increase again until it approaches the steady state value~(for the finite-temperate thermal state with $p_+<1/2$). In contrast, if the initial state is the ground state, the purity
decreases monotonously, both for thermal bath and dephasing noise.

\section*{1D spinless Fermion chain subjected to a linear potential}

\subsection{Dependence of the entropy on temperature}

\begin{figure}[htb]
  \centering
  \includegraphics[width=0.58\columnwidth]{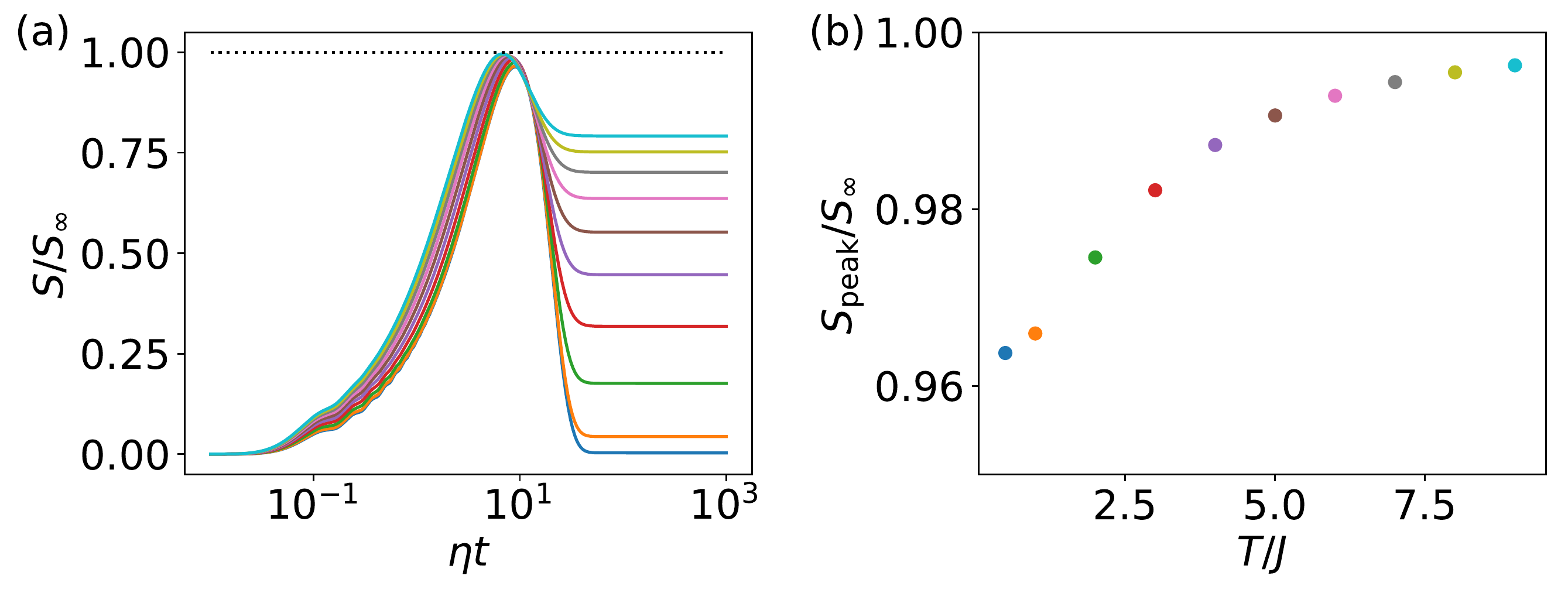}
  \caption{(a)~Time evolution of the von Neumann entropy $S$
for a half-filling spinless Fermion chain whose dynamics is governed by Eq.~\eqref{heat}. The dotted line marks the maximum entropy $S_{\infty}$ for the maximally mixed state. The initial state is a Fock state with the left side of the chain occupied. (b)~Peak entropy $S_{\rm peak}$ as a function of the temperature of the thermal bath. The parameters are $M=8$, $V=J$, $r=4J$, $\eta=0.1J$.}\label{Tvec}
\end{figure}

Figure~\ref{Tvec}(a) shows the dynamics of the von Neumann entropy $S$
for a half-filling spinless Fermion chain described by Hamiltonian \eqref{Ham} in the main text coupled to
a thermal bath at different temperatures. 
The peak entropy during the evolution shows a
weak dependence on the temperature, with a larger entropy at a higher temperature, as shown in Fig.~\ref{Tvec}(b).

\subsection{Dependence of the entropy on the initial state}

\begin{figure}[htb]
  \centering
  \includegraphics[width=0.58\columnwidth]{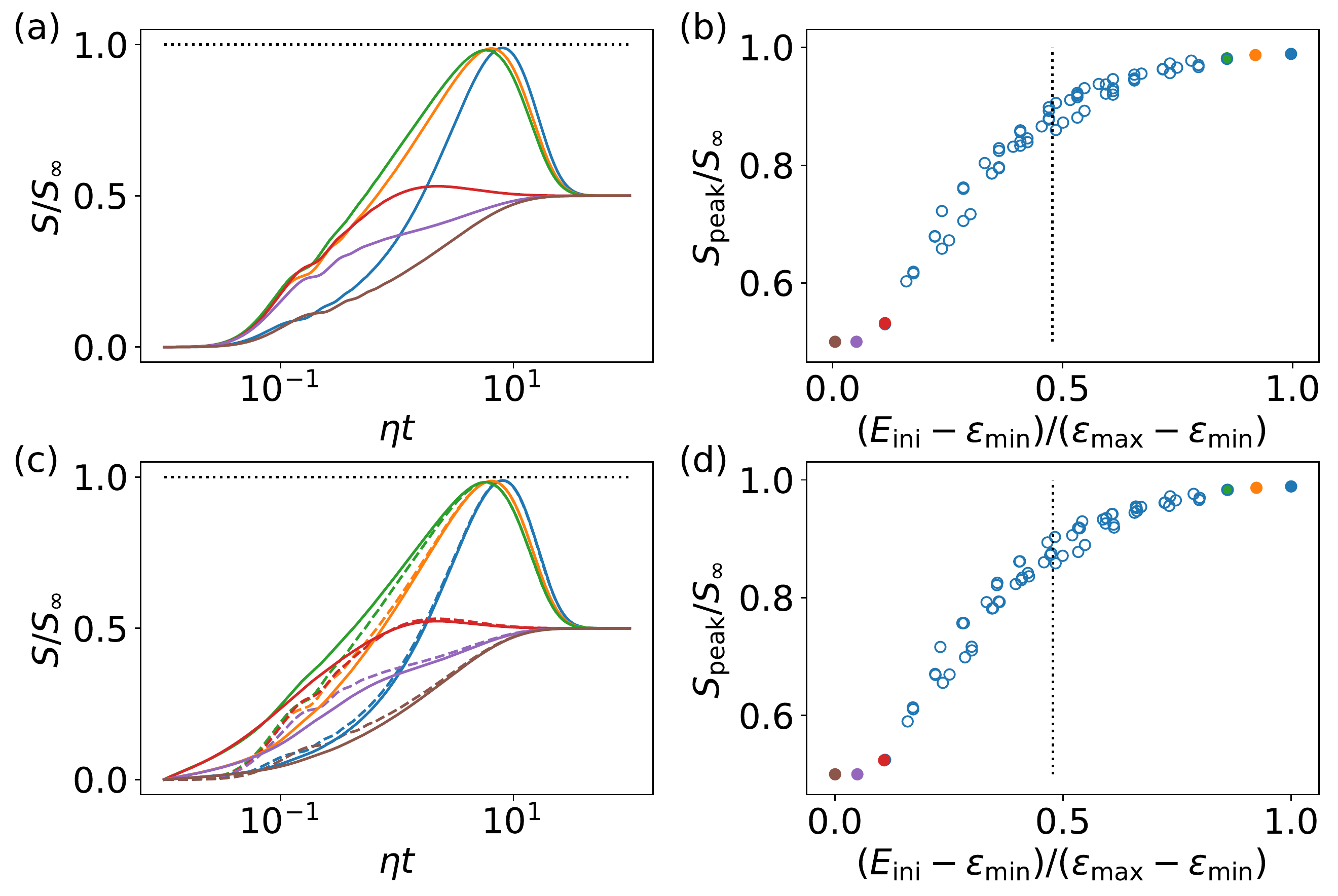}
  \caption{(a),(c)~Time evolution  of the von Neumann entropy $S$ for a half-filling spinless Fermion chain coupled to a thermal bath described by Eq.~\eqref{heat}.
    The dotted line marks the maximum entropy $S_{\infty}$ for the maximally mixed state.
    The temperature of the thermal bath is chosen such that $S_T = S_{\infty}/2$.
    (b),(d)~Peak entropy $S_{\rm peak}$~(normalized by $S_{\infty}$) as a function of the averaged energy of the
    initial state $E_{\rm ini}$ (scaled between 0, for the ground-state energy $\varepsilon_\text{min}$,
    and 1, for the energy $\varepsilon_\text{max}$ of the most excited state).
     The dotted line marks the average energy of the system, which is also the mean energy of the maximally mixed state (infinity-temperature state), $E_\infty=\mathrm{tr}\{\rho_\infty H\}$.
     {The initial states for (a), (b) are Fock states, and the initial states for (c), (d) are many-body energy eigenstates. The dashed lines in (c) are the results of (a).}
     The parameters are $M=8$, $V=J$, $r=4J$, $\eta=0.1J$.}\label{init}
\end{figure}

In Fig.~\ref{init} we compare the time evolution of the entropy for different initial states. While (a) and (b) correspond to initial Fock states, (c) and (d) capture the dynamics starting from many-body energy eigenstates. While for the former scenario the initial density matrix possesses off-diagonal elements in energy representation, this is not the case the for tlatter one. Thus, the fact that both scenarios show almost identical behavior, indicates that the off-diagonal elements of the initial Fock states have decayed before reaching the peak entropy.

\subsection{Universal behavior of mean occupation in real space}

\begin{figure}[!htbp]
    \centering
    {\includegraphics[width=0.95\columnwidth]{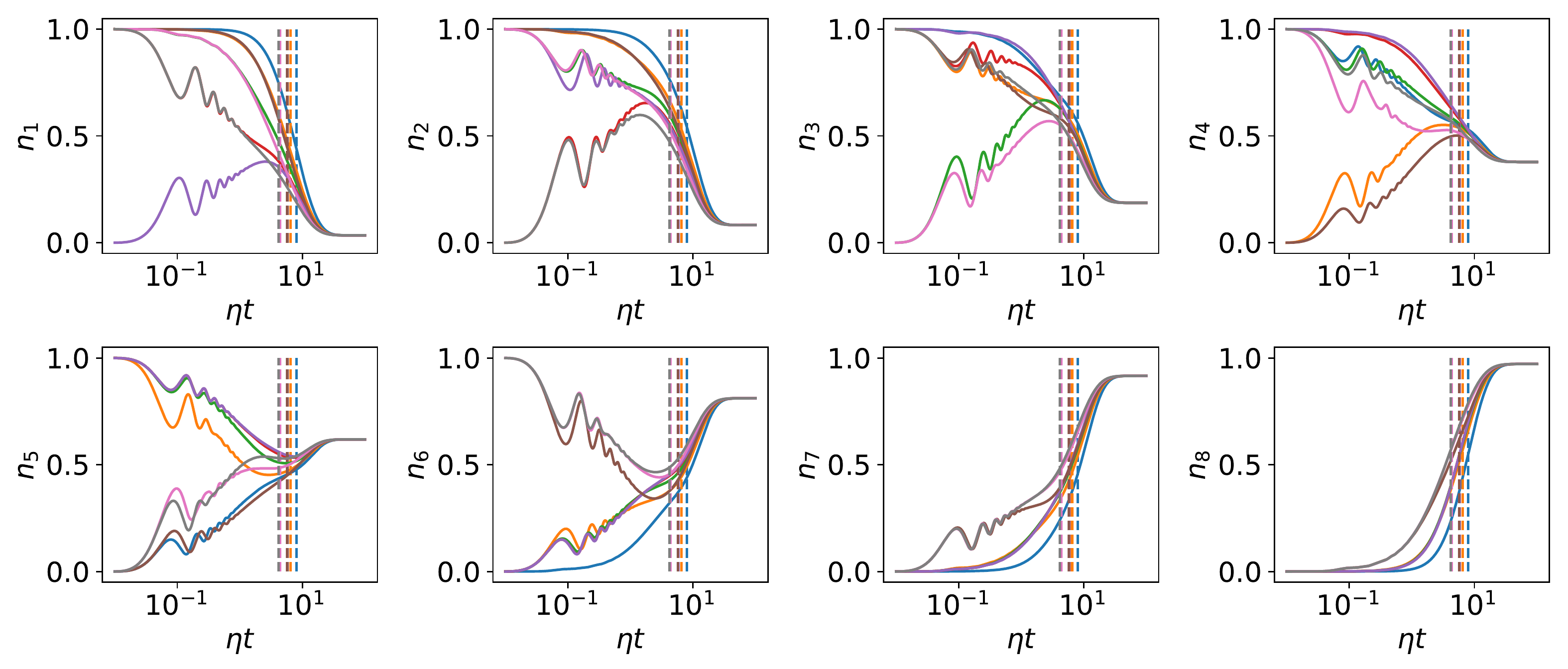}}
    \caption{Time evolution of the mean occupation in real space $\langle n_i\rangle$ for a half-filling Fermion chain [with Hamiltonian~\eqref{Ham}] coupled to
    a thermal bath described by Eq.~\eqref{heat}.
     The vertical dashed lines marks the time when the entropy gets maximal.
    The initial states are $7$ Fock states with the highest energies.
    The parameters are $M=8$, $r=4J$, $V=J$, $\eta=0.1J$.
    }\label{nt}
\end{figure}

\begin{figure}[!htbp]
    \centering
    {\includegraphics[width=0.95\columnwidth]{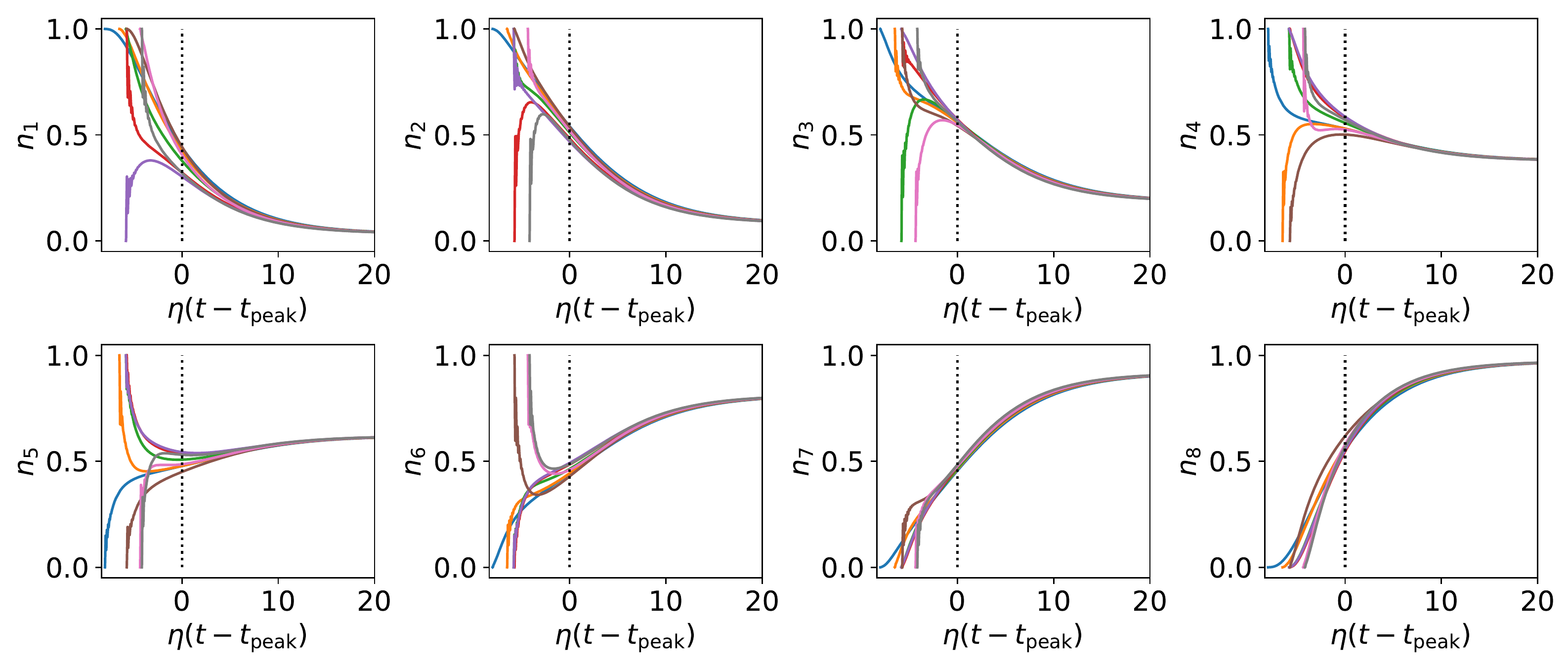}}
    \caption{The same as Fig.~\ref{nt} except that the time $t$ is shifted by the time to get the peak entropy $t_{\rm peak}$. The black dashed line marks $t=t_{\rm peak}$.
    }\label{ntv}
\end{figure}

Figure \ref{nt} shows time evolution of the mean occupation in real space $\langle n_i\rangle$ for a half-filling chain under a linear potential described by Hamiltonian~\eqref{Ham} starting from different initial Fock states.
The dashed lines denote the time $t_{\rm peak}$ when the peak entropy is reached.
By plotting
the evolution of the site occupation relative to
the time $t_{\rm peak}$,
the different curves clearly converge near $t = t_{\rm peak}$, as shown in Fig.~\ref{ntv}.

\begin{figure}
  \centering
  \includegraphics[width=0.6\columnwidth]{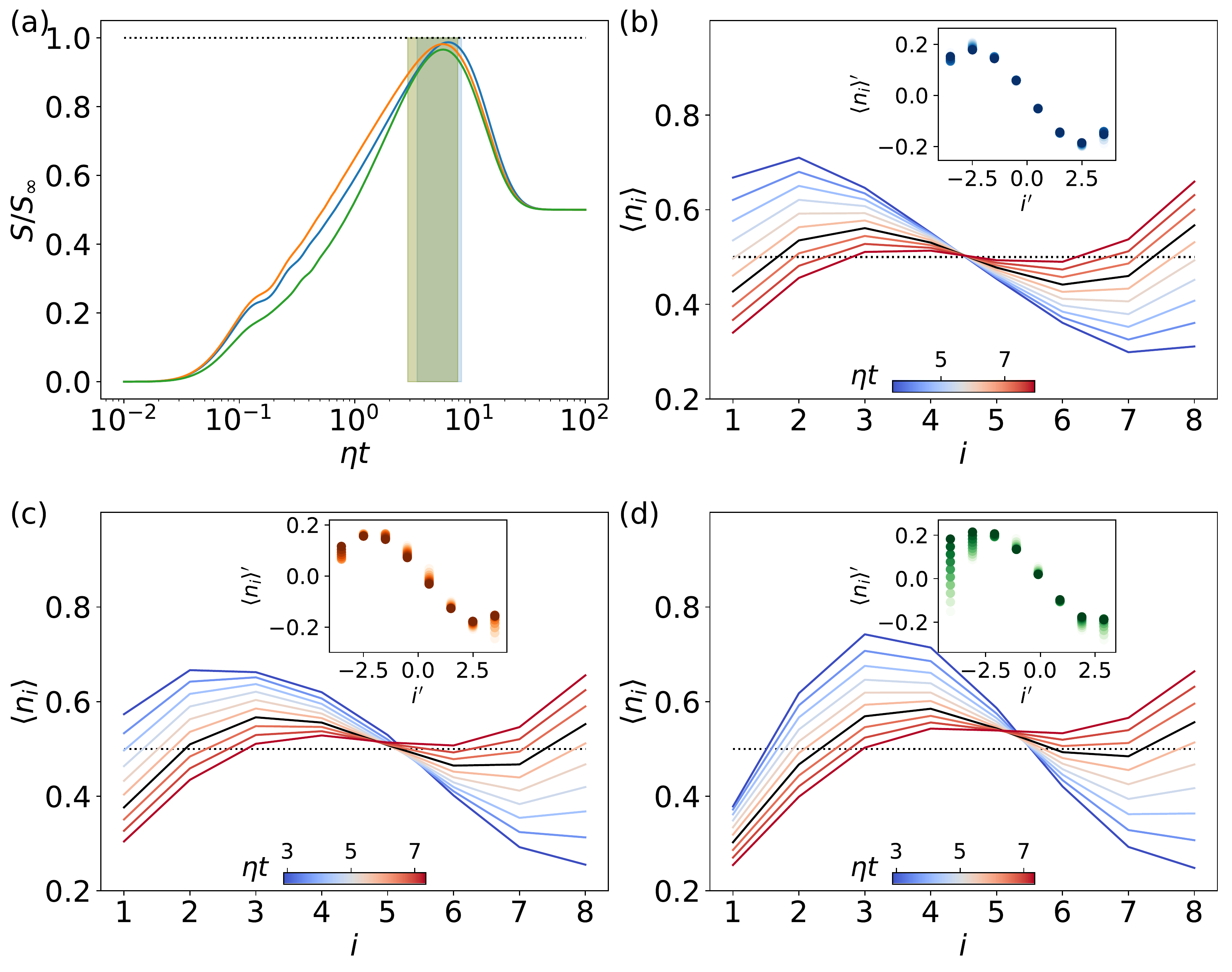}\\
  \caption{(a)~Time evolution of the von Neumann entropy $S$ for a half-filling spinless Fermion chain [with Hamiltonian~\eqref{Ham}] coupled to a thermal bath
    described by Eq.~\eqref{heat} starting from three different initial Fock states.
 (b)-(d) Mean occupation in real space for the time window marked by shading area in (a).  The inset shows the collapse of the curves by rotating them by an angle proportional to the corresponding time. The axis in the inset is $i^\prime = (i-i_0)\cos[\chi(t-t_0)] + (\langle n_i\rangle - y_0)\sin[\chi(t-t_0)]$, and $\langle n_i\rangle^\prime = (\langle n_i\rangle - y_0)\cos[\chi(t-t_0)] - (i-i_0)\sin[\chi(t-t_0)]$ with the parameters $x_0$, $y_0$, $t_0$ and $\chi$ adjusted to get the optimal collapse.  The parameters are $M=8$, $V=J$, $r=4J$, $\eta=0.1J$, $T$ is chosen to make $S_T=S_{\infty}/2$.}\label{ni}
\end{figure}

In Fig.~\ref{ni}, we show the mean occupation in real space starting from three different initial Fock states [(b)-(d)] for the time window marked by shading area in (a).
The density profiles at different times in the vicinity of peak entropy are found to share similar shape. As shown in the inset, by rotating these curves
by an angle proportional to the corresponding time, they collapse onto each other.

\subsection{Results for larger systems}
\subsubsection{Pauli rate equation}
\begin{figure}[H]
  \centering
  \includegraphics[width=0.9\columnwidth]{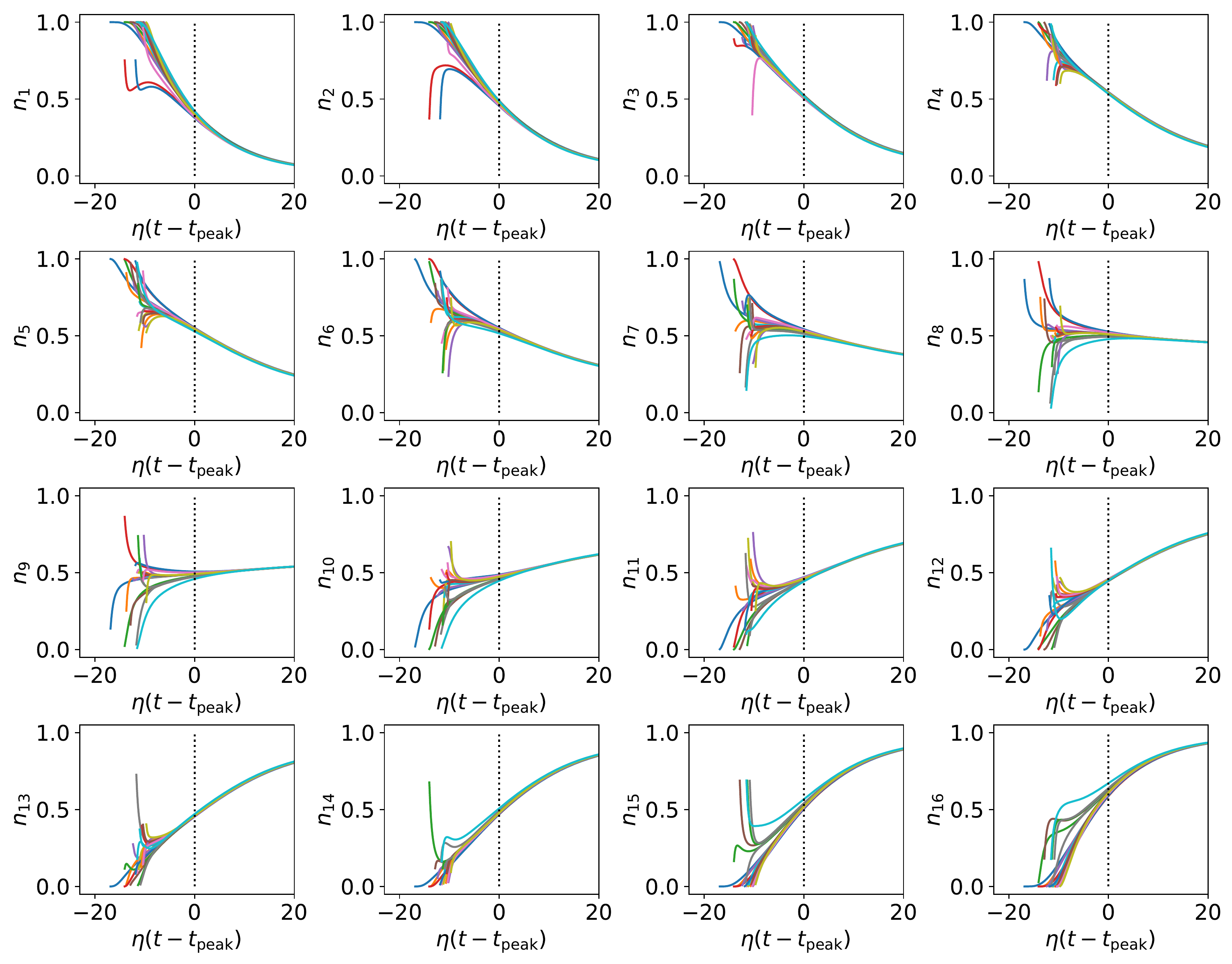}
  \caption{ Time evolution of the mean occupation in real space $\langle n_i\rangle$ starting from different initial states with high energies. The density matrix is approximated by $\rho=\sum_k p_k |k\rangle \langle k|$ with $p_k$ governed by Eq.~\eqref{rate_eq}. The parameters are $M=16$, $V=J$, $r=4J$.}\label{n_rate_eq}
\end{figure}

According to the Lindblad master equation introduced in the main text, the dynamics of the diagonal and off-diagonal elements of the density matrix are decoupled.
By neglecting the off-diagonal elements~(which allows us to study larger system), the density matrix is given by $\rho=\sum_k p_k |k\rangle \langle k|$, with the diagonal elements
$p_k$ governed by the Pauli rate equation
\begin{equation}\label{rate_eq}
{\dot p_k} = \eta \sum_q {\left( {{R_{kq}}{p_q} - {R_{qk}}{p_k}} \right)}.
\end{equation}
Figure~\ref{n_rate_eq} shows time evolution of the mean occupation in real space $\langle n_i\rangle$ for a half-filling Fermion chain with $M=16$ sites by solving Eq.~\eqref{rate_eq}.
We find similar universal dynamics as in small system~[see Fig.~\ref{ntv}].

\subsubsection{Kinetic theory}
The equations of motion for the mean occupations in single-particle eigenstates are
given by
\begin{equation}
\langle {\dot n}_k \rangle = \sum_q {\left[{\tilde R}_{kq} \langle (1-n_k)n_q \rangle - {\tilde R}_{qk}\langle (1-n_q)n_k \rangle\right]},
\end{equation}
where ${\tilde R}_{kq}=\sum_i |\psi_{ik}|^2|\psi_{iq}|^2$ and $\psi_{iq}$ is the single particle eigenstate.
In order to obtain a closed set of equations in terms of the
mean occupations, we employ the mean-field approximation,
\begin{equation}
\langle n_k n_q \rangle \simeq \langle n_k\rangle \langle n_q \rangle,
\end{equation}
for $k \ne q$. Then we obtain a set of nonlinear kinetic equations of motion
\begin{equation}
\langle {\dot n}_k \rangle = \sum_q {\left[\left({\tilde R}_{kq} \langle n_q \rangle - {\tilde R}_{qk}\langle n_k \rangle\right)
-\left({\tilde R}_{kq}-{\tilde R}_{qk}\right)\langle n_k \rangle \langle n_q\rangle\right]},
\end{equation}
which can be solved numerically.

\begin{figure}[htb]
  \centering
  \includegraphics[width=0.9\columnwidth]{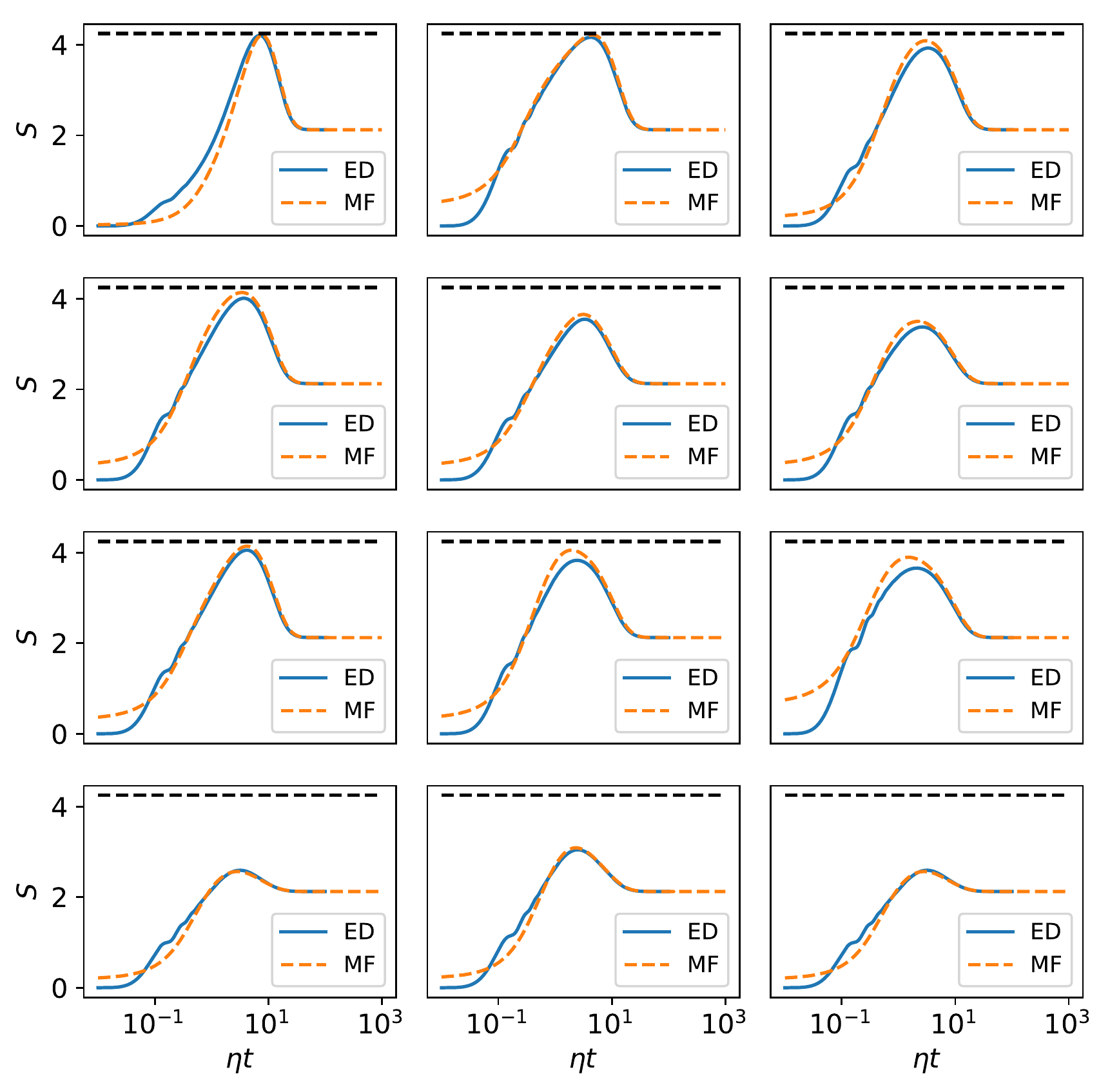}
  \caption{Comparison of time evolution of entropy from numerically solving Eq.~\eqref{heat} (solid lines)
  and from kinetic theory~(dashed lines) for different initial Fock states. The parameters are $M=8$, $V=0$, $r=4J$. }\label{ED_MF_S}
\end{figure}

\begin{figure}[H]
  \centering
  \includegraphics[width=0.9\columnwidth]{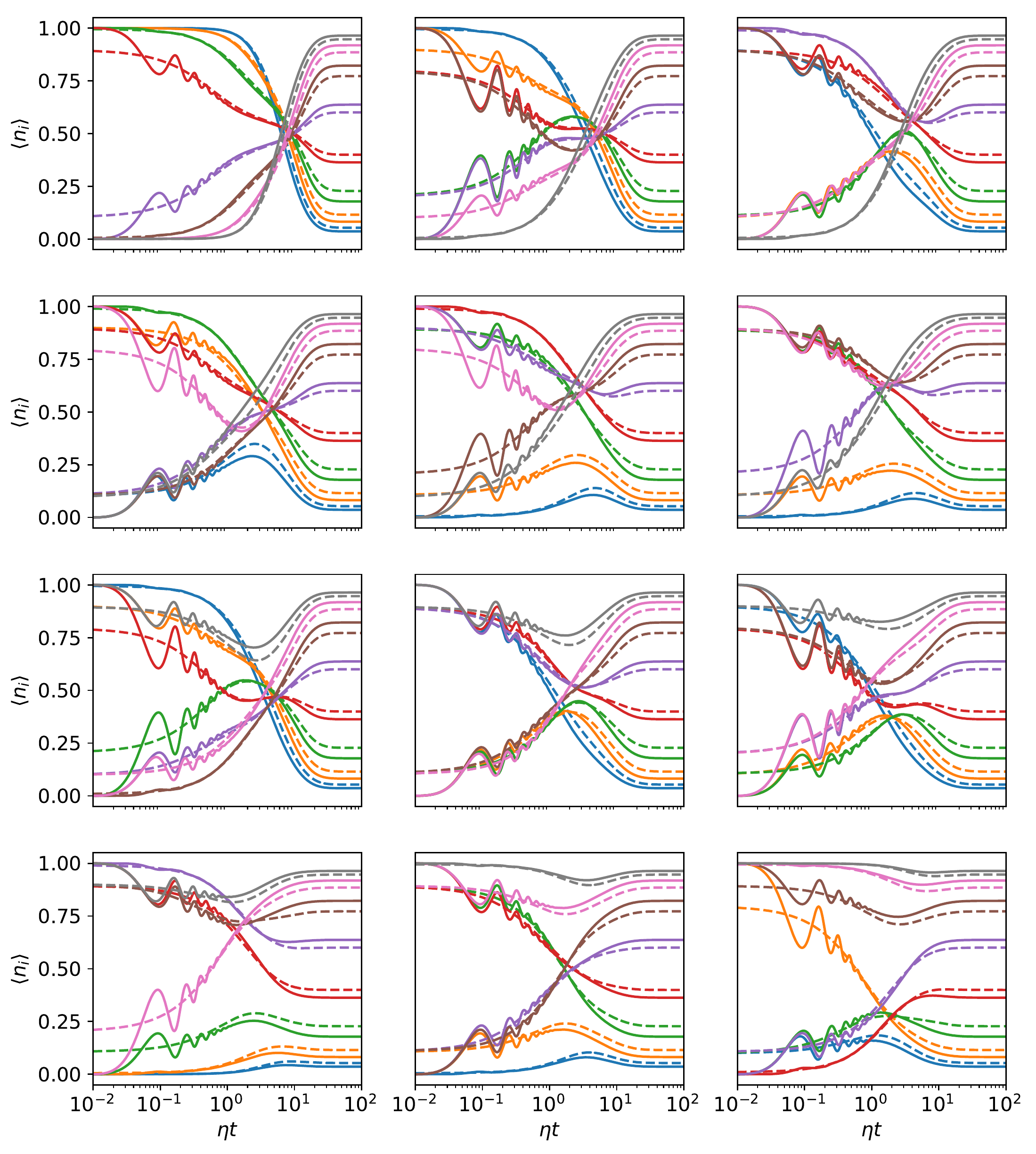}
  \caption{ Comparison of time evolution of the mean occupation  in real space $\langle n_i\rangle$ from numerically solving Eq.~\eqref{heat} (solid lines)
  and from kinetic theory~(dashed lines) for different initial Fock states. The parameters are $M=8$, $V=0$, $r=4J$.}\label{ED_MF_n}
\end{figure}

\begin{figure}[H]
  \centering
  \includegraphics[width=0.9\columnwidth]{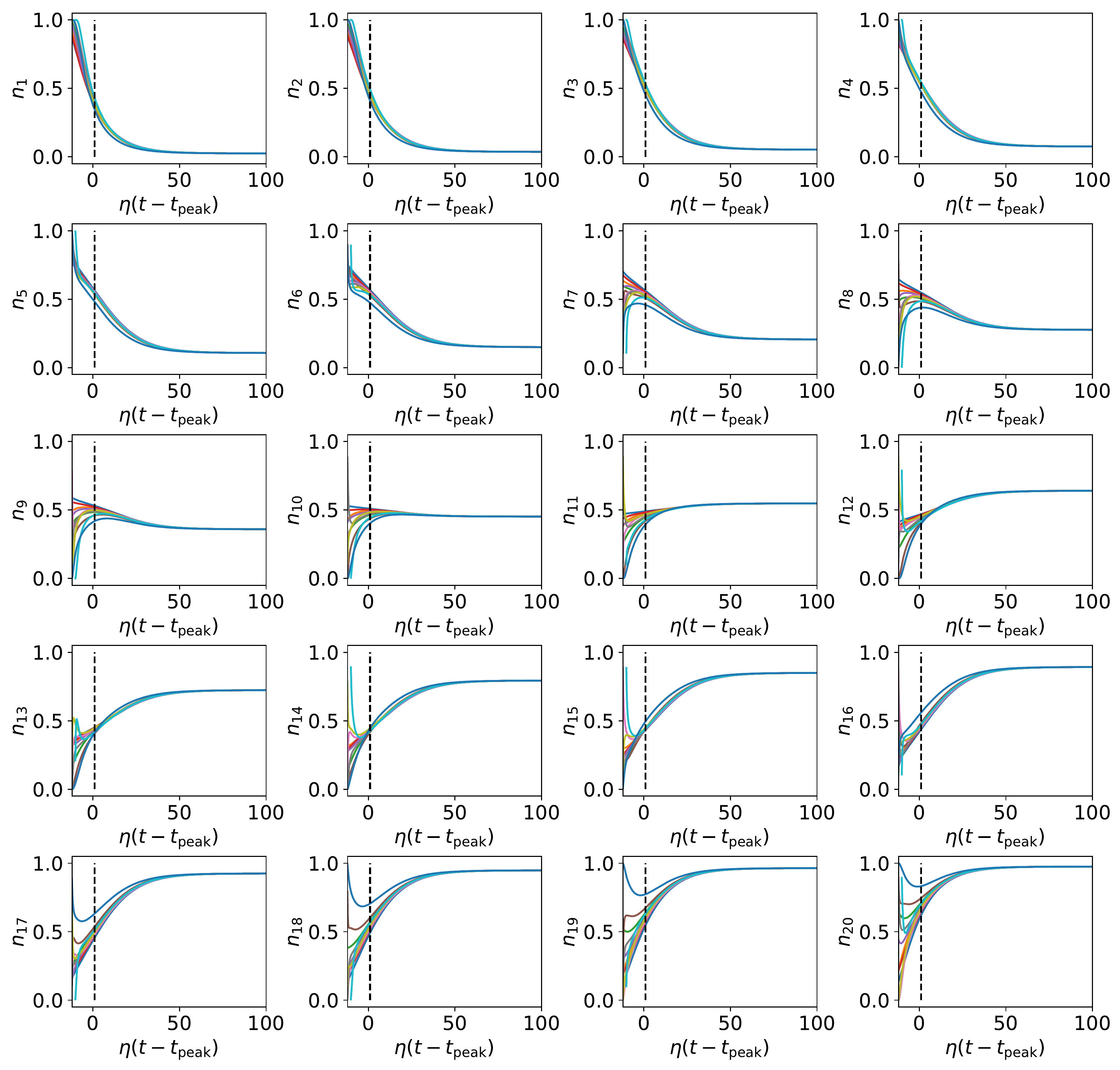}
  \caption{Time evolution of the mean occupation  in real space $\langle n_i\rangle$ from kinetic theory for a half-filling chain under a linear potential described by Hamiltonian~\eqref{Ham}.
     The vertical dashed lines marks the time of reaching peak entropy.
    The initial states are different Fock states with high energies.
    The parameters are $M=20$, $r=4J$, $\eta=0.1J$.}\label{nmf}
\end{figure}

The mean-field approximation is equivalent to a Gaussian ansatz for the density matrix
\begin{equation}\label{mf}
\rho = \frac{1}{\cal Z}\exp(-\sum_k \alpha_k n_k),
\end{equation}
where ${\cal Z}$ is the partition function and
$\alpha_k$ is determined by the $M$ mean occupations
$\langle n_k \rangle$, as $\alpha_k = \ln(\langle n_k\rangle^{-1}-1)$.
Figure~\ref{ED_MF_S} compares the time evolution of entropy for a half-filling chain of $M=8$ sites from exact diagonalization (solid lines)
  and from mean-field theory~(dashed lines) for different initial Fock states.
  Figure~\ref{ED_MF_n} compares the time evolution of the corresponding
  mean occupation in real space.
  The deviations that are visible in the steady state result from the fact that Eq.~\eqref{mf} corresponds to the grand canonical ensemble, rather than to a canonical Gibbs state.
Figure~\ref{nmf} shows the time evolution of the mean occupation in real space for a half-filling chain of $M=20$ sites from mean-field theory starting from
different initial Fock states.

\section*{Disordered Potential}

\begin{figure}[!htbp]
  \centering
\includegraphics[width=0.6\columnwidth]{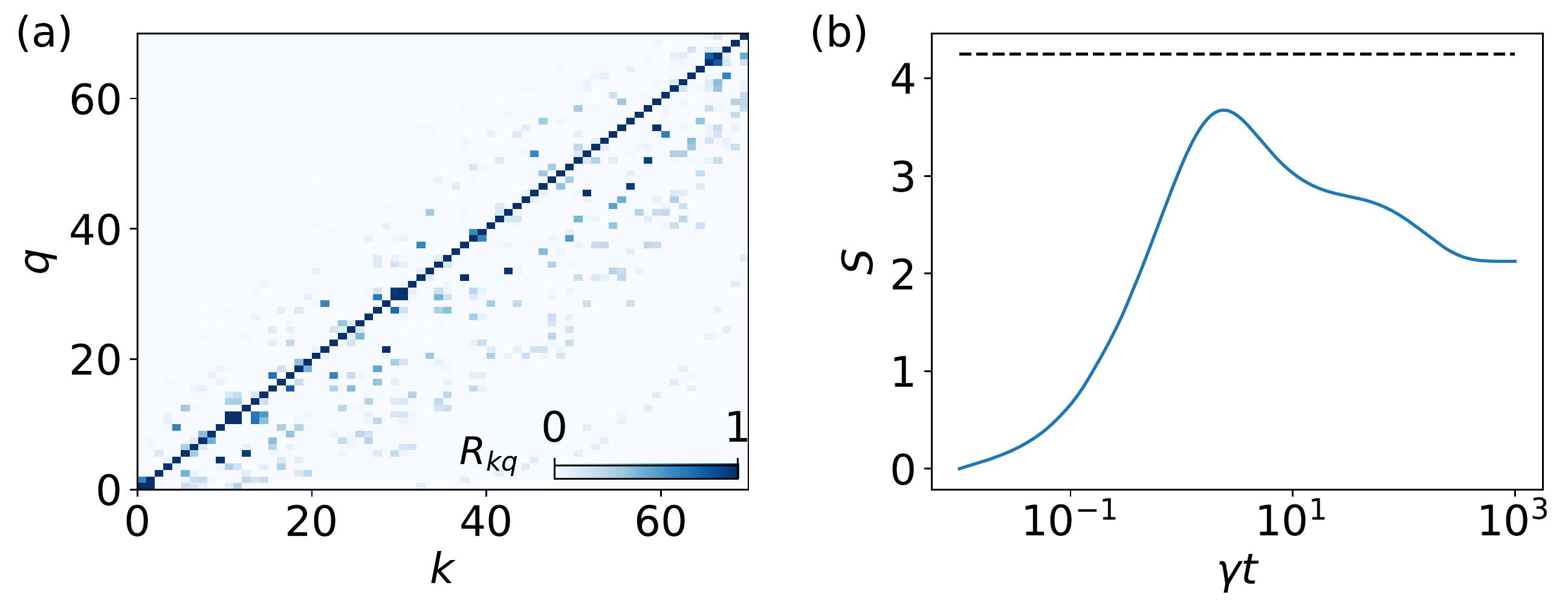}

  \caption{(a)~Rate matrix for a Fermi-Hubbard chain under a disordered potential with disorder strength $W=20J$ coupled to a thermal bath. (b)~
  Time evolution of entropy. The initial state is the highest excited state. The results are for one disorder realization. The parameters are $M=8$, $V=J$, $\eta=0.1J$, $T$ is chosen to make $S_T=S_{\infty}/2$.}\label{disorder}
\end{figure}

Fig.~\ref{disorder}(a) shows the rate matrix for a Fermi-Hubbard chain under a disordered potential with random on-site energies uniformly drawn from the interval $[-W,W]$ coupled to a thermal bath.
It has a non-local structure, which is different from that of the Stark model in the main text~[see Fig.~2(b)].
Such a rate matrix can not give rise to a close-to-maximum peak entropy, as shown in Fig.~\ref{disorder}(b).

\section*{Optimized potential model}

\begin{figure}[!htbp]
  \centering
\includegraphics[width=0.3\columnwidth]{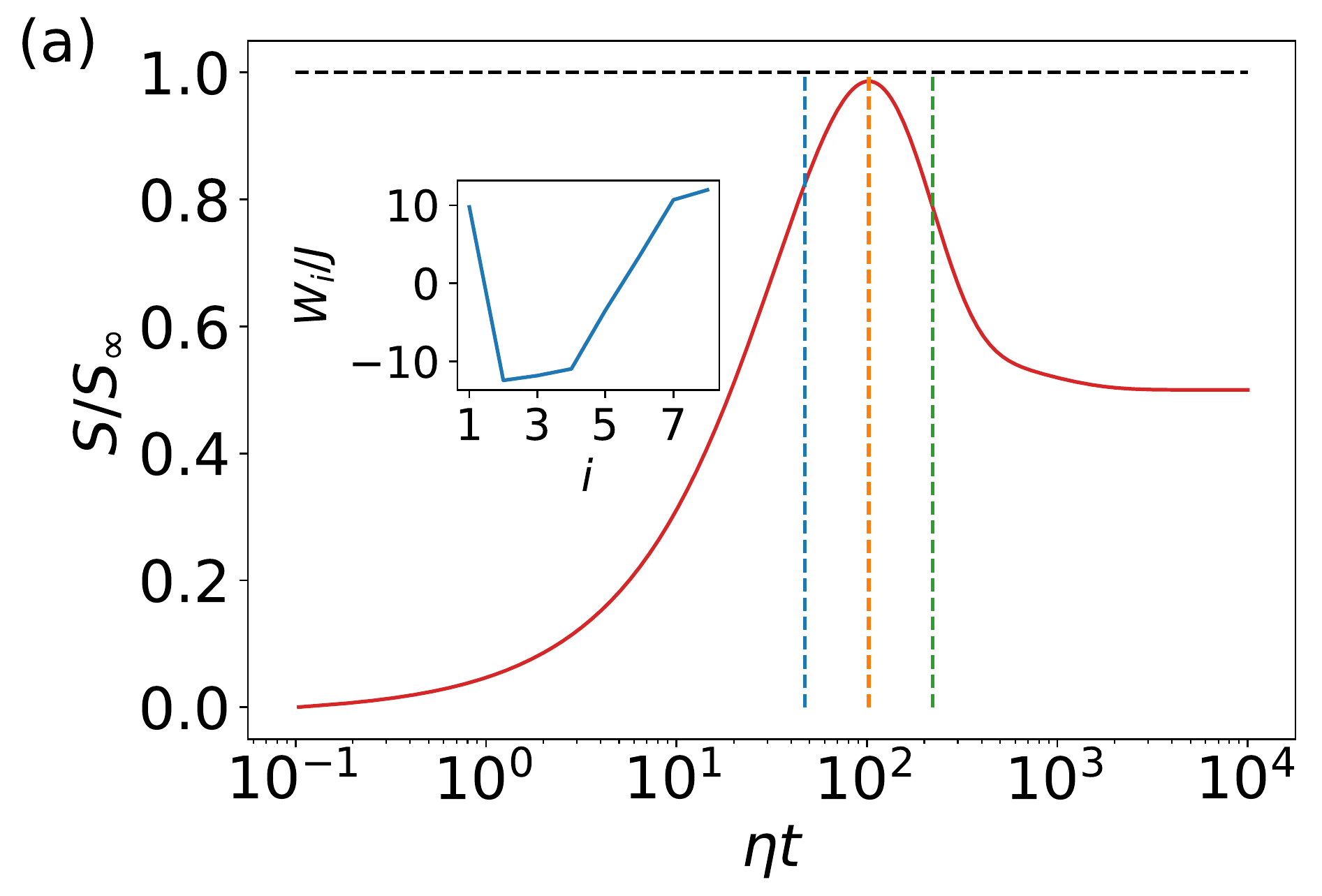}
\includegraphics[width=0.3\columnwidth]{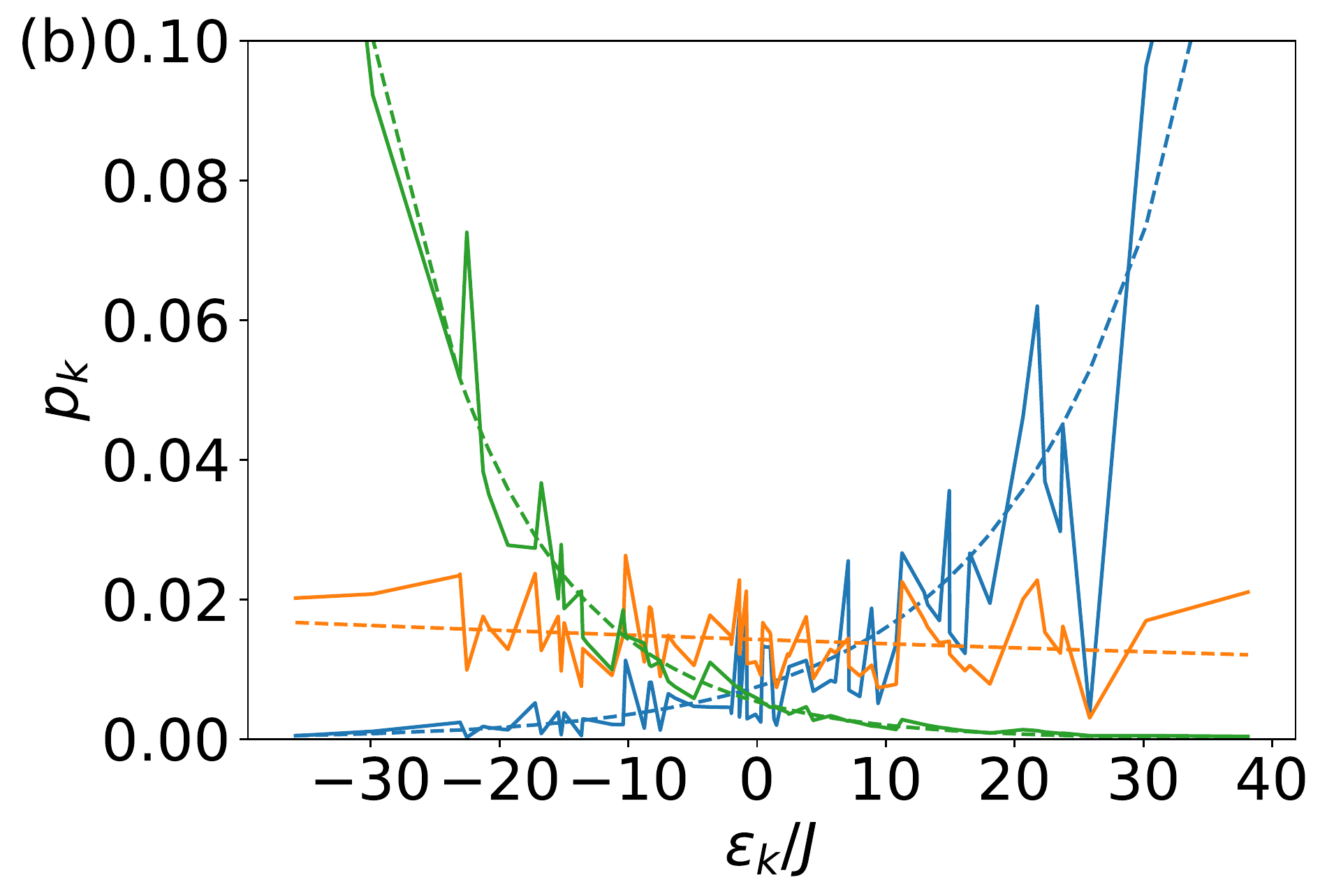}

  \caption{(a)~Time evolution of entropy $S =- \sum_k p_k \log(p_k)$ for a half-filling chain (described by Hamiltonian~\eqref{Ham} with the on-site potential $W_i$ shown in the inset) coupled to a thermal bath. The dynamics of $p_k$ is governed by rate equation~\eqref{rate_eq}. (b)~Distribution of $p_k$ at three moments marked in (a) are shown in solid lines,
    which is close to the distribution of the corresponding
    thermal states with the same average energy shown in dashed lines. The parameters are $M=8$, $V=J$, $\eta=0.1J$, $T$ is chosen to make $S_T=S_{\infty}/2$.}\label{OP-thermal1}
\end{figure}

Figure~\ref{OP-thermal1} shows time evolution of entropy $S =- \sum_k p_k \log(p_k)$ for a half-filling chain (described by Hamiltonian~\eqref{Ham} with the on-site potential $W_i$ shown in the inset) coupled to a thermal bath. The dynamics of $p_k$ is governed by rate equation. The distribution $p_k$ at the three different times marked by vertical lines in (a) is shown by solid lines in (b). It is found to be close to a thermal distribution, analogous to the Stark model~[see Fig.~1(f) in the main text].

%
%

\section*{Simplified models}
\begin{figure}[!htbp]
  \centering
  \includegraphics[width=0.7\columnwidth]{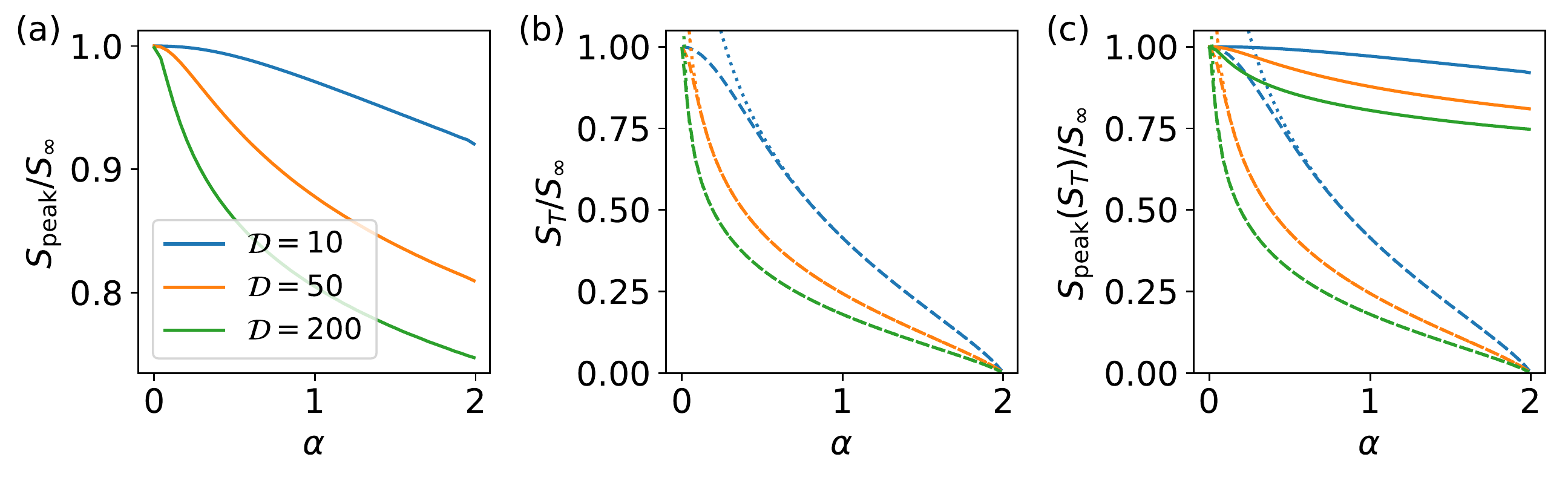}
  \includegraphics[width=0.7\columnwidth]{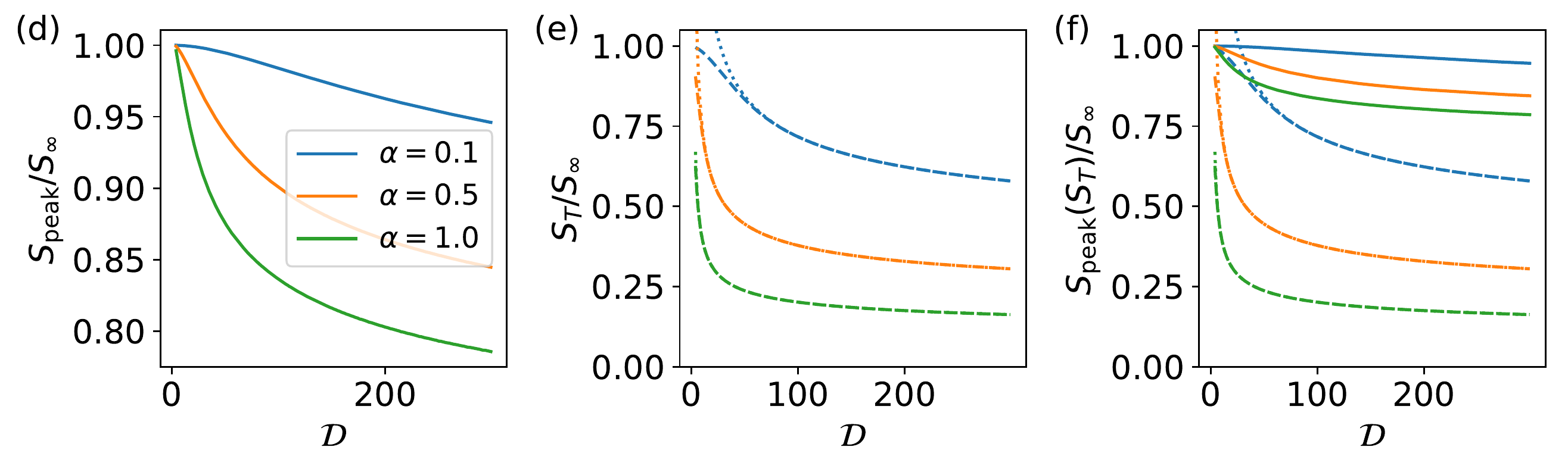}
  \caption{The peak entropy ($S_{\rm peak}$) and the thermal entropy ($S_{\rm T}$)~(normalized by the maximum entropy $S_{\infty}=\log({\cal D})$)
as a function of (a)-(c) $\alpha \equiv \delta R/\bar R$ and (d)-(f) ${\cal D}$.
  The entropy is calculated as $S =- \sum_k p_k \log(p_k)$, with $p_k$ governed by rate equation~\eqref{rate_eq}. The initial state is $p_k=\delta_{k,{\cal D}-1}$ for $0 \le k \le {\cal D}-1$.
The dotted lines are from the expression
  $S_{T} = \frac{x}{x-1}\log(x)-\log(1-x)$ with $x=(2-\alpha)/(2+\alpha)$.}\label{Sam}
\end{figure}

Figure~\ref{Sam} shows the normalized peak entropy $S_{\rm peak}$ and thermal entropy $S_{ T}$
for the simplified model where $p_k$ is governed by $\dot p_k = \eta(\bar R \nabla^2 p_k + \delta R \nabla p_k)$
starting from the highest excited state. In (a)-(c), the dependence of $S_{\rm peak}$ and $S_{T}$  on $\alpha=\delta R/\bar R$ for three different
system dimensionality ${\cal D}$ is shown. In (d)-(f), the dependence of $S_{\rm peak}$ and $S_{T}$ on ${\cal D}$ for three different
${\alpha}$ is shown. The results indicate the existence of a non-trivial parameter regime with both $S_\text{peak}\approx S_\infty$ and $S_T$ well below $S_{\infty}$.

\begin{figure}[!htbp]
  \centering
  \includegraphics[width=0.8\columnwidth]{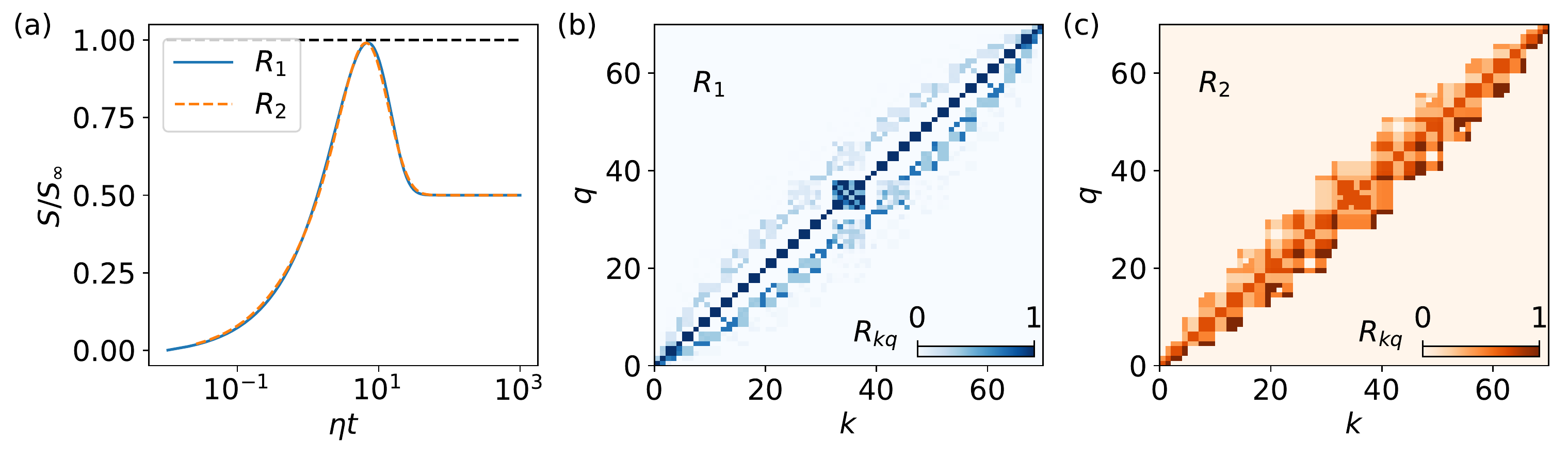}
  \includegraphics[width=0.8\columnwidth]{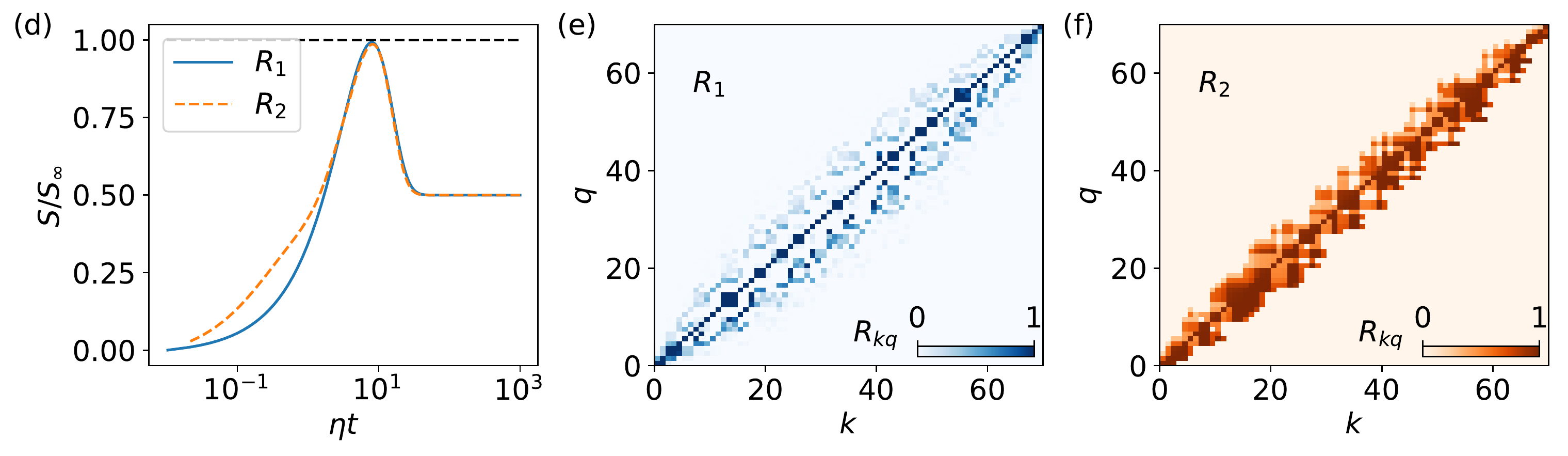}
  \caption{
(a)~Time evolution of the entropy $S =- \sum_k p_k \log(p_k)$ with $p_k$ governed by rate equation~\eqref{rate_eq} under two different rate matrices shown in (b) and (c).
  The initial state is the highest excited state. The parameters are $M=8$, $V=0$, $r=4J$, $\eta=0.1J$, $\epsilon=0.15$.
  (d)-(f) show the corresponding results for interacting system with $V=J$. The fitting parameter is $\epsilon=0.22$ in this case.}\label{rate_texture}
\end{figure}

%


In Fig.~\ref{rate_texture}(a) we show the time evolution of entropy $S =- \sum_k p_k \log(p_k)$
with $p_k$ governed by rate equation under two different rate matrices shown in (b) and (c).
The former~($R_1$) is the real rate matrix of the noninteracting Stark model. It depends on both the bath correlation
function and the overlap of wave functions, i.e., $R_{kq}=\pi v_{kq} g(\varepsilon_k-\varepsilon_q)$
with $v_{kq} = \sum_i{|\langle k|n_i |q\rangle|^2}$, which endows it a complex texture.
The rate matrix $R_2$ used in (c) is obtained from $R_1$ by replacing $v_{kq}$ by the binary values $\epsilon$ or $0$, depending on whether $|v_{kq}|>0.1$ or $<0.1$, respectively.
As shown in Fig.~\ref{rate_texture}(a), the entropies for these two rate matrices are found to be almost the same.
In (d)-(f), we show the corresponding results for interacting system with $V=J$.

\end{document}